\documentclass[aip,reprint,twoside,a4paper,floatfix]{revtex4-1}

\usepackage{amsmath}
\usepackage{xcolor}
\usepackage{amsfonts}
\usepackage{amssymb}
\usepackage{graphicx}
\usepackage{dcolumn}
\usepackage{bm}

\usepackage[utf8]{inputenc}
\usepackage[T1]{fontenc}

\usepackage{mathptmx}
\usepackage{etoolbox}
\usepackage{siunitx} 
\DeclareSIUnit{\year}{a}
\usepackage{float} 

\usepackage{datetime2}

\usepackage{datetime2}
\usepackage{fancyhdr}
\makeatletter
\def\@email#1#2{%
 \endgroup
 \patchcmd{\titleblock@produce}
  {\frontmatter@RRAPformat}
  {\frontmatter@RRAPformat{\produce@RRAP{*#1\href{mailto:#2}{#2}}}\frontmatter@RRAPformat}
  {}{}
}%
\makeatother

\newcommand{\cf}{{\textit{cf.\@}}}

\newcommand{\eq}[1]{(\ref{eq:#1})}
\newcommand{\Eq}[1]{Eq.~\eq{#1}}
\newcommand{\EQ}[1]{Equation~\eq{#1}}
\newcommand{\Eqs}[1]{Eqs.~\eq{#1}}

\newcommand{\fig}[1]{\ref{fig:#1}}
\newcommand{\Fig}[1]{Fig.~\fig{#1}}
\newcommand{\Figs}[1]{Figs.~\fig{#1}}
\newcommand{\FIG}[1]{Figure~\fig{#1}}

\newcommand{\sect}[1]{\ref{sec:#1}}
\newcommand{\Sect}[1]{Sect.~\sect{#1}}
\newcommand{\Secs}[1]{Sects.~\sect{#1}}

\newcommand{\app}[1]{\ref{app:#1}}
\newcommand{\App}[1]{App.~\app{#1}}

\newcommand{\rmd}{\ensuremath{\mathrm{d}}}
\newcommand{\rme}{\ensuremath{\mathrm{e}}}

\begin{document}

\date{\DTMnow}

\title{Constraining safe and unsafe overshoots in saddle-node bifurcations}

\author{Elias Enache}
\affiliation{\text{Institute for Theoretical Physics, University of Leipzig, Germany}}

\author{Oleksandr Kozak}
\affiliation{\text{Institute for Theoretical Physics, University of Leipzig, Germany}}

\author{Nico Wunderling}
\homepage{ORCID: \href{https://orcid.org/0000-0002-3566-323X}{0000-0002-3566-323X}}
\affiliation{FutureLab Earth Resilience in the Anthropocene, Potsdam Institute for Climate Impact Research (PIK), Member of the Leibniz Association, Potsdam, Germany.}
\affiliation{Stockholm Resilience Centre, Stockholm University, Stockholm, Sweden.}

\author{J\"urgen Vollmer}
\homepage{ORCID: \href{https://orcid.org/0000-0002-8135-1544}{0000-0002-8135-1544}}
\affiliation{\text{Institute for Theoretical Physics, University of Leipzig, Germany}}

\date{\today}

\begin{abstract}
  We consider a dynamical system undergoing a saddle-node bifurcation
  with an explicitly time dependent parameter~$p(t)$.
  The combined dynamics can be considered as a dynamical systems 
  where $p$ is a slowly evolving parameter.
  Here, we investigate settings where the parameter features an overshoot.
  It crosses the bifurcation threshold for some finite duration $t_e$ and up to an amplitude $R$,
  before it returns to its initial value.
  We denote the overshoot as safe when the dynamical system returns to its initial state.
  Otherwise, one encounters runaway trajectories (tipping), and the overshoot is unsafe.
  For shallow overshoots (small $R$) safe and unsafe overshoots are discriminated by an inverse square-root border, $t_e \propto R^{-1/2}$,
  as reported in earlier literature.
  However, for larger overshoots we here establish a crossover to another power law with an exponent 
  that depends on the asymptotics of $p(t)$. 
  For overshoots with a finite support we find that $t_e \propto R^{-1}$,
  and we provide examples for overshoots with exponents in the range $[-1, -1/2]$.
  All results are substantiated by numerical simulations,
  and it is discussed how the analytic and numeric results pave the way towards improved risks assessments separating safe from unsafe overshoots in climate, ecology and nonlinear dynamics.
\end{abstract}


\maketitle

\begin{quotation}
  Tipping points are critical thresholds in dynamical systems
  at which small perturbations may lead to a qualitative and possibly irreversible change of the dynamical system's state.\cite{1984HorsthemkeLefever,2009SchefferBacompteBrock-EtAl,2013Lenton,2015BrummittBarnettDSouza,2018FeudelPisarchikShowalter,2023DatserisRossiWagemakers}
  Often these tipping points are crossed only up to a certain amplitude $H$ and for some finite duration $T$,
  before parameters return to their sub-threshold values.
  In geology this applies for heavy rainfalls inducing land slides.\cite{1980Caine,1993LarsenSimon,2017ChaeParkCataniSimoniBerti}
  In economy this is a common scenario for bank stress tests.\cite{2008MayLevinSugihara}
  In neural dynamics it is a core idea in the criticality hypothesis modeling the brain computational properties.\cite{2021Gross}
  In the Earth's climate system the tipping points for melting the large ice sheets on Greenland or Antarctica may be crossed at least temporarily due to ongoing global warming.\cite{Ritchie2,wunderling2023global}
  In all cases it is vital to identify the boundary between safe overshoots, where the system returns to its initial state without tipping,
  and unsafe overshoots, where tipping events are triggered.
  Earlier literature\cite{Ritchie} reported an inverse square-root law $T \sim H^{-1/2}$ as boundary between safe and unsave overshoots.
  Here, we introduce an dimensionless overshoot amplitude $H$ and duration $T$,
  and we establish that this result applies for shallow overshoots $H \simeq 1$.
  For larger overshoot amplitudes $H \gtrsim 1$ there is a crossover to a steeper power-law decay  
  with exponents in the range between $-1/2$ to $-1$.
  We provide upper bounds for parameter profiles that decay with different powers, and
  by numerical work we show that these bounds are sharp.
  Hence, the safe region in parameter space is substantially smaller than expected,
  with severe consequences for the risks assessments of separating safe from unsafe overshoots.
\end{quotation}

\section{Introduction}

Tipping emerges when an asymptotic state of a dynamical system disappears in a saddle-node bifurcation.\cite{2011Kuehn,2012AshwinWieczorekVitoloCox,2013Lenton,2015BrummittBarnettDSouza,2018FeudelPisarchikShowalter}
We will characterize the state of the system by the scalar variable $x(t)$,
and assume that its equilibrium state $x_\infty$ changes continuously upon changing a control parameter $r$ for values $r<0$.
At $r=0$ the dynamical system undergoes a saddle-node bifurcation
where $x_\infty$ fuses with an unstable fixed point:
For $r<0$ there is a region of initial conditions that converge towards $x_\infty$,
and for $r>0$ these initial conditions will run away to a different attractor.
Here we are interested in situations where the control parameter is varied in time $r(t)$:
For late and early times it approaches $R_d < 0$,
at time $t=0$ it takes its maximum value $R > 0$,
and it takes positive values for a time interval of duration $t_e$.
We denote this behavior as an overshoot of duration $t_e$ and amplitude $R$. 
When the overshoot is sufficiently shallow and short, i.e.~when $R$ and $t_e$ are both small,
the system will return to its initial state.
We denote this as a \emph{safe overshoot}.
An overshoot where the system runs away to another attractor will be denoted as \emph{unsafe overshoot}.

Overshoots have been discussed in the field of nonlinear systems and bifurcation theory over recent years.\cite{Ritchie,Ritchie2}
The main question for general systems is, whether we observe a return to the original state after the control parameter falls below the critical value again.
This question may be relevant for various applications in finance, politics, or ecology.\cite{brummitt2015coupled,rocha2018cascading}
Moreover, recently overshoots received particular attention in climate science,\cite{drouet2021net,riahi2021cost,rogelj2019new}
because there are several subsystems of the Earth that may exhibit threshold behavior.

\begin{figure}
\centering
\includegraphics{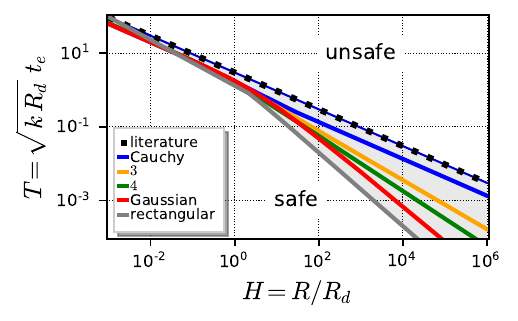}
\caption{ \label{fig:TH_comparison}
  The border dividing the parameter space spanned by the dimensionless overshoot amplitude $H$ and overshoot duration $T$ (both introduced in \Eq{HT}) into a safe region (no tipping) and an unsafe region (tipping).
  The different line refer to different shapes of the overshoot with the same amplitude and duration.
  The uppermost, dotted black line corresponds to the results from literature\cite{Ritchie} and follows a power law.
  Numerical results from the present article for an overshoot following a Gaussian shape are represented by the red solid line,
and those for distributions decaying to their asymptotic values with power laws with exponents $2$ (Cauchy distribution),
$3$ and $4$ by blue, orange and green solid lines respectively.
The solid gray line provides the boundary for a discontinuous, piecewise constant dependence of $r(t)$.
} 
\end{figure}

Based on a multiple time scale analysis \cite{1987BenderOrszag-book,2011Kuehn}
\citeauthor{Ritchie}\cite{Ritchie} established
that the border separating safe and unsafe overshoots takes the form of a power law,
$R \: t_e^2 = $const,
provided that $R$ is small, $t_e$ is large, and that the overshoot takes the form of a parabola.
This dependence is shown by the uppermost, black dotted line in \Fig{TH_comparison}. 
In the present paper we extend this result to all values of $R$ and arbitrary form of the overshoot
(other lines in \Fig{TH_comparison}):
\\
For $R \lesssim R_d$ we recover the power law suggested in the literature.
Different shapes of $r(t)$ only slightly modify the prefactor of the power law.
\\
For $R \gtrsim R_d$ the boundary changes qualitatively.
It crosses over to a steeper power law with an exponent that depends on the asymptotic decay of $r(t)$ towards $R_d$. 
We determine the corresponding power, and potential logarithmic corrections.

The paper is organized as follows.
In \Sect{criterium-jump} we determine the boundary of the stable region for a piecewise constant overshoot trajectories.
Subsequently, we use this analytical result to establish upper and lower bounds for the boundaries
of systems where the overshoots deviates from $R_d$ only for a finite time (\Sect{bounds4finite-support})
and where they take a general unimodal shape (\Sect{bounds4unimodal}).
In \Sect{Gaussian} we discuss the bounds for overshoots with a Gaussian shape,
as previously considered in the literature\cite{Ritchie}
and \Sect{kappa} deals with overshoots with power-law tails.
In the discussion, \Sect{discussion},
we revisit the earlier findings,\cite{Ritchie}
and discuss the relevance of our findings for climate models.
Key findings are summarized in \Sect{conclusion}.

\section{ Solution for rectangular overshoots }
\label{sec:criterium-jump}

We consider a system described by a variable $x(t)$
with a dynamics close to a saddle-node bifurcation upon varying a parameter, $r$.
Following the theory of normal forms\cite{1983GuckenheimerHolmes} we require
that the bifurcation arises at $r=0$, 
and we approximate the dynamics as
\begin{equation} \label{eq:EOM}
    \dot{x} = k\, x^2 + r(t), \qquad \text{with constant \ } k > 0 \, .
\end{equation}
We do not absorb $k$ into the time scale
because the natural time scale of the dynamics 
is provided by the duration $t_e$ where $r(t)$ takes positive values.
For $r<0$ the dynamics has  an unstable fixed point at $\sqrt{r/k} > 0$, 
and a stable fixed point at $-\sqrt{r/k}$.
At $r=0$ the two fixed points collide.
For $r > 0$ all trajectories will grow without bounds.
We denote this as tipping.

We are interested in scenarios where the parameter $r$ takes the time dependence of an overshoot.
It starts off and decays to a value of $-R_d < 0$
\begin{subequations}
\begin{align} 
  \lim_{|t| \to \infty} r(t) &= -R_d  < 0 \, ,
\end{align}
it takes positive values in the interval $t \in ] -t_e/2 , t_e/2 [$,
\begin{align} 
  |t|   <  \frac{t_e}{2}  \qquad&\Rightarrow \qquad  r(t) > 0  \, ,\\
  |t| \geq \frac{t_e}{2}  \qquad&\Rightarrow \qquad  r(t) \leq 0 \, ,
\end{align}
and it takes a maximum value of $R$,
\begin{align} 
  \max_t r(t) &= R > 0 \, .
\end{align}
\end{subequations}
Hence, the control parameter exceeds the bifurcation threshold for a time span $t_e$,
and it reaches into the unstable parameter domain till a maximum value of $R$. 
To gain a qualitative understanding about the implications of an overshoot we first consider the scenario
or a rectangular overshoot.
Subsequently, we will generalize the findings to other functional dependences, $r(t)$.

\begin{figure*}
\[
  \includegraphics{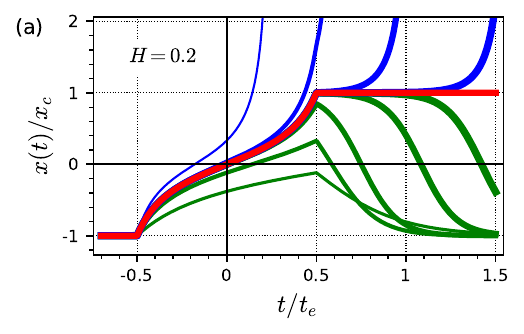} \qquad
  \includegraphics{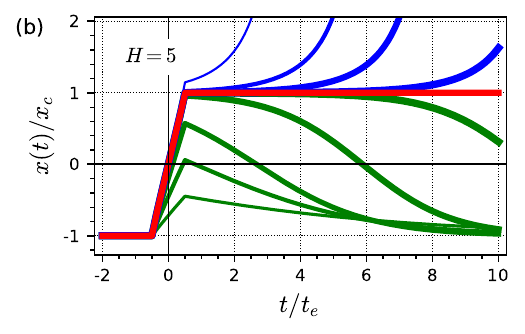} 
\]
\caption{ \label{fig:jump-trajectories}
  The different curves represent solutions for $x(t)$ during the overshoot ($|t|< t_e/2$)
  and afterwards ($t>t_e/2$), as outlined in \Eqs{solutiontan} and~\eqref{eq:threecases}, respectively.
  The stable fixed point $x=-x_c$ serves as a common initial condition before the overshoot starts at $t=-t_e/2$.
  The blue curves represent the divergent solutions with $x(t_e/2) > x_c$.
  The green curves represent the solutions with $x(t_e/2) < x_c$ that converge back to the stable fixed point at $-x_c$.
  The thick red line corresponds to $x(t_e/2) = +x_c$, the constant solution at the unstable fixed point.
  \\
  (a) Trajectories for $H=0.2$ with
  green lines for $T = 2, \; 4, \; 5, \; 5.14, \; 5.144$,
  blue lines for $T = 5.14415, \; 5.15, \; 5.5, \; 8$,
  and the red line $T_c = (2 /\sqrt{0.2}) \; \text{atan}( 1/\sqrt{0.2} ) \simeq 5.144128$.
  (b) Trajectories for $H=5$ with
  green lines for $T = 0.1, \; 0.2, \; 0.3, \; 0.37, \; 0.376$,
  blue lines for $T = 0.3762, \; 0.377, \; 0.38, \; 0.4$,
  and the red line $T_c = (2 /\sqrt{5}) \; \text{atan}( 1/\sqrt{5} ) \simeq 0.3761373$.
  The thickness of the lines decreases with increasing distance of $T$ from the $T_c$.
} 
\end{figure*}

For rectangular overshoots 
\begin{align} \label{eq:rectangular-overshoot}
 r(t) = \begin{cases}
             \:\;\; R \qquad & \text{for \ } |t|  <   \frac{t_e}{2},\\[1mm]
             -R_d            & \text{for \ } |t| \geq \frac{t_e}{2},
            \end{cases}
\end{align}
the dynamics~\Eq{EOM} can be integrated analytically.
At early times we start off at the stable fixed point $-x_c = \sqrt{R_d/k}$,
\begin{align*} 
  t < - \frac{t_e}{2} \: : \qquad
  \frac{ x(t) }{ x_c } = -1 \, ,
\end{align*}
and for $-t_e/2 < t < t_e/2$ the control parameter $r(t)$ takes the constant value of $R$.
In that case~\Eq{EOM} can be solved by separation of variables.
For the initial condition $x(-t_e/2) = -x_c$ one obtains,
\begin{widetext}
\begin{align}  \label{eq:solutiontan}
  -\frac{t_e}{2} \leq t \leq \frac{t_e}{2} \: : \qquad
  \frac{ x(t) }{ x_c }
  = \sqrt\frac{R}{R_d} \;
    \tan\!\left[ \sqrt{kR} \: \left( t + \frac{t_e}{2} \right) - \text{atan}\sqrt\frac{R_d}{R} \right]
  = \sqrt H \;
    \tan\!\left[ T\, \sqrt H \: \left( \frac{t}{t_e} + \frac{1}{2} \right) - \text{atan}\frac{1}{ \sqrt H } \right] \, .           
\end{align}
\end{widetext}
Here the dimensionless parameters
\begin{align}   \label{eq:HT}
  H = \frac{R}{R_d}
  \qquad \text{and}\qquad 
  T = \sqrt{k \, R_d} \;\; t_e
\end{align}
characterize the strength and the duration of the overshoot.
Some characteristic trajectories are shown in \Fig{jump-trajectories}.

\begin{figure*}
\[
  \includegraphics{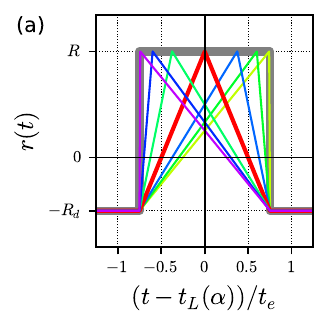} \rule{-3mm}{0mm}
  \includegraphics{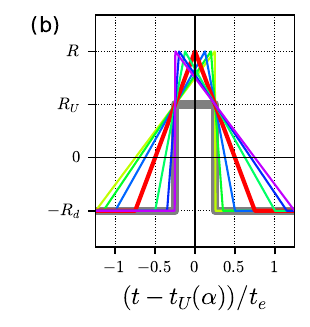} \quad
  \includegraphics{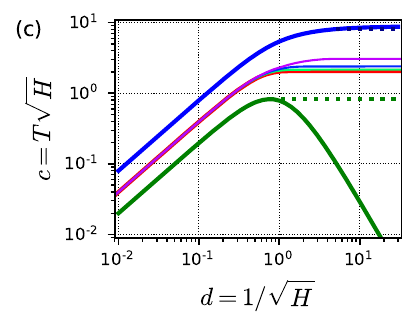}
\]
\caption{\label{fig:triangular-bounds}
  Panels (a) and (b) show different triangular-shaped, i.e.~piecewise-linear profiles $r(t)$,
  where $\alpha$ denotes the position of their respective maxima.
  They are all bounded from above (panel a) by a profile
  that takes the value $R$ on an interval of duration $(1+d^2)\,t_e$ and $-R_d$ outside the interval.
  They are all bounded from below (panel b) by a profile
  that takes the value $R_U$ on an interval of duration $(1-R_U/R) \, t_e$ and $-R_d$ outside the interval.
  The profiles in panels (a) and (b) are shifted by $t_L(\alpha)$ and $t_U(\alpha)$, respectively,
  in order to emphasize that $\alpha$ only induces a trivial time shift of the piecewise-constant bounds.
  According to \Eqs{bound-triangle} these bounds provide the boundaries
  indicated by the green and the blue line in panel (c).
  The numerical values for the position of the stability boundary are marked by lines with colors matching those in the other two panels.
  The dotted green line indicates that for sufficiently large~$d$ one can improve the lower bound by considering
  overshoots that do not reach down to $-R_d$, as in panel (a), but rather take a larger value $-R_d < -R_L < 0$.
  Full details are provided in the main text.
}
\end{figure*}

The subsequent evolution of the trajectory, for $t > t_e/2$, is described by \Eq{EOM} with a constant parameter $r=-R_d$.
The asymptotics of $x(t)$ depends on the value of $x( t_e/2 )$.
\\
When $x( t_e/2 )$ lies to the left of the unstable fixed point, $x( t_e/2 ) < x_c = \sqrt{R_d/k}$,
the variable $x(t)$ will relax back to the stable fixed point at $-x_c$.
\\
When  $x( t_e/2 )$ lies to the right of the unstable fixed point, $x( t_e/2 ) > x_c$,
it will run away to infinity.
\\
For $x( t_e/2 ) = x_c$, the variable $x(t)$ will reside at this unstable fixed point:
\begin{widetext}
  \begin{align}
    t > \frac{t_e}{2} \; \land \; x(t_e/2) > x_c \: : \qquad
    \frac{ x(t) }{ x_c }
    &= -\text{cotanh}\!\left[ T \: \left( \frac{t}{t_e} - \frac{1}{2} \right) - \text{acotanh}\: \frac{ x(t_e/2) }{x_c} \right] \, ,    
    \nonumber \\ \label{eq:threecases}
    t > \frac{t_e}{2} \; \land \; x(t_e/2) = x_c \: : \qquad
    \frac{ x(t) }{ x_c } &= 1 \, ,
    \\
    t > \frac{t_e}{2} \; \land \; x(t_e/2) < x_c \: : \qquad
    \frac{ x(t) }{ x_c }
    &= -\text{tanh}\!\left[ T \: \left( \frac{t}{t_e} - \frac{1}{2} \right) - \text{atanh}\: \frac{ x(t_e/2) }{x_c} \right] \, .
      \nonumber
  \end{align}
\end{widetext}
For a given value of $H$ the values of $x(t)/x_c$ increase monotonically with $T$.
Therefore, the stability boundary is provided by the trajectories that terminate at the unstable critical point,
\begin{align*} 
  1 &= \frac{ x(t_e/2) }{ x_c }
  =  \sqrt H \;
  \tan\!\left[ T\,\sqrt H \: - \text{atan}\frac{1}{\sqrt H} \right]
\end{align*}
which provides the stability boundary for rectangular overshoots
\begin{align} \label{eq:jump-boundary}
  T_R &= \frac{ 2 }{ \sqrt H } \;\; \text{atan}\frac{1}{\sqrt H}
\end{align}
that is marked by the gray line in \Fig{TH_comparison} (i.e.,~the lowermost line).
It shows a crossover from a decay $T \simeq \pi / \sqrt{H}$ for small values of $H$,
where the arcus tangens takes values close to $\pi/2$,
to a decay $T \simeq 2 / H$ for large values of $H$,
where the arcus tangent approaches the identity.

\Eq{jump-boundary} reveals how the crossover in $T$ from an $H^{-1/2}$ decay to $H^{-1}$ emerges for $H \gtrsim 1$
(cf.~\Fig{TH_comparison}).
Moreover, it also suggest that the latter regime may conveniently be studied in terms of the dimensionless variables
\begin{subequations} \label{eq:boundary-cd}
\begin{align}   \label{eq:def-cd}
  c = T \: \sqrt H = \sqrt{ k R } \: t_e
  \quad\text{and}\quad
  d = \frac{1}{\sqrt{H}} = \sqrt{\frac{R_d}{R}}
\end{align}
such that \Eq{jump-boundary} takes the form
\begin{align}   \label{eq:c-jump-boundary}
  c_R = 2 \;\: \text{atan}( d ) \, .
\end{align}
\end{subequations}

In the following we show that the crossover is generic.
It is observed for all parameter profiles
where the order parameter decays to its asymptotic values faster than $t^{-2}$.

\section{ Bounds for the stability border }
\label{sec:bounds}

In order to derive bounds for the stability boundary we compare the trajectories $x_1(t)$ and $x_2(t)$ 
for the overshoots $p_1(t)$ and $p_2(t)$.
Their difference evolves as
\begin{align}  \label{eq:monotonicity-condition}
  \frac{\rmd}{\rmd t} ( x_2 - x_1 )
  = k \: (x_2 + x_1 ) \; (x_2-x_1) + ( p_2 - p_1 ) \, .
\end{align}
For small values of $x_2-x_1$
the right hand side of this equation takes the form of an exponential for $p_2 = p_1$,
and it is strictly positive for $p_2 > p_1$.

Let the trajectories start at their respective stable fixed points.
Then $p_2 \geq p_1$ implies that $x_2 - x_1$ starts off positive semi-definite,
and that $x_2$ becomes strictly larger than $x_1$ as soon as $p_2$ takes larger values than $x_1$. 
Combining this monotonicity condition with the result of \Sect{criterium-jump} provides bounds for the position of the stability borders.
For overshoots of triangular shape this is demonstrated in \Fig{triangular-bounds}.

\subsection{Overshoots with finite support}
\label{sec:bounds4finite-support}

We say that an overshoot has a finite support when there is a constant $\Delta > 1$ with
\begin{align*} 
  \left\lvert t \right\rvert > \Delta \; t_e
  \quad \Rightarrow \quad
  r(t) = -R_d \, .
\end{align*}

A lower bound to the stability border is then provided by the rectangular overshoot
\begin{align*} 
  r(t) = \left\{ \begin{aligned}
    -R_d &\quad\text{for}\quad \left\lvert t \right\rvert > \Delta \; t_e \, , \\
    R    &\quad\text{else.}
  \end{aligned} \right.
\end{align*}
Hence, \Eqs{jump-boundary} and \eqref{eq:boundary-cd} apply up to the substitution $t_e \rightarrow \Delta \, t_e$,
i.e.~with $d_L = d = \sqrt{R_d/R}$ and $c_L = \Delta \: c = \sqrt{ k R } \: \Delta\: t_e$.
Consequently, we obtain the following lower bound $c_L$
for the stability boundary of overshoots with finite support,
\begin{subequations} \label{eq:bound-triangle}
\begin{align} \label{eq:bound-triangle-lower1}
  c_L = \frac{c_R}{\Delta} = \frac{2}{\Delta} \; \text{atan}( d ) \, .
\end{align}
For triangular-shaped overshoots we have $\Delta = 1+d^2$. 
The corresponding line is plotted as a solid green line in \Fig{triangular-bounds}(c).

The upper bound is provided by a rectangular overshoot where $R_U$ takes the role of $R$
such that the duration of the overshoot decreases by a factor of $\delta < 1$.
Moreover, its amplitude is decreased by a factor of $R_U/R < 1$.
With $T_U = \delta\: T$ and $H_U = (R_U/R) \: H$, and \Eqs{def-cd} and \eqref{eq:jump-boundary}
one obtains an upper bound $c_U$
for the stability boundary of overshoots with finite support,
\begin{align} \label{eq:bound-triangle-upper1}
  c_U = \frac{c_R}{\delta \: \sqrt{R_U/R}}
  = \frac{2}{\delta \: \sqrt{R_U/R}}\; \text{atan}\frac{ d }{ \sqrt{ R_U/R} }  \, .
\end{align}%
\end{subequations}%
The optimal, i.e.~lowermost upper bound is obtained by taking into account the relation between $\delta$ and $R_U$,
and minimizing the right hand side of \Eq{bound-triangle-upper1}
for different choices of $\delta$ in the range $0 < \delta < 1$.

For triangular-shaped overshoots we have $\delta = 1-R_U/R$ such that 
\begin{align} \label{eq:bound-triangle-upper}
  c_U = \frac{2}{\delta \: \sqrt{1-\delta}} \; \text{atan} \frac{d}{\sqrt{1-\delta}} \, .
\end{align}%
For small values of $d$ the choice $\delta=1/2$ is optimal, as one rapidly checks after introducing
$\text{atan}(d/\sqrt{1-\delta}) \simeq d/\sqrt{1-\delta}$.
The solid blue line in \Fig{triangular-bounds}(c) provides this boundary. 
For large $d$ one rather has $\text{atan}(d/\sqrt{\Delta}) \simeq \pi/2$, 
and the value $\delta=2/3$ provides slightly smaller bound.
It is provided by a dotted blue line.

For small $d$ the lower bound provided by \Eq{bound-triangle-lower1} provides a good idea about the boundary.
However, for $d \gtrsim 1$ it is rapidly decreasing while one would rather need a constant asymptotic value. 
A better lower bound can obtained by adopting a rectangular overshoot profile with a lower value $0 > -R_L = -R\: (\Delta-1) > -R_d$
with $1 < \Delta < 1+d^2$.
The smallest width that provides an upper bound will then be $\Delta \: t_e$,
and we obtain
\begin{align} \nonumber
  c_L &= \max_{\Delta > 1} \frac{2}{\Delta} \; \text{atan}( \sqrt{\Delta-1} )
  \\[2mm]
  \label{eq:bound-triangle-lower}
  &= \left\{
  \begin{aligned}
    &0.8239\dots  &&\text{for} \quad d > \sqrt{ \Delta_c-1 } \simeq 0.7654 \, , \\
    &\frac{2}{1+d^2} \; \text{atan}( d ) \quad  &&\text{else} \, .
  \end{aligned}
  \right .
\end{align}
This bound takes it maximum value for $\Delta_c \geq 1.5858$.
This bound is provided by the dashed green line in \Fig{triangular-bounds}(c).
For $d < \sqrt{ \Delta_c-1 } \simeq 0.7654$ this value can not be realized
because the support of the overshoot has a width of $\Delta_\text{max} = 1+d^2$.
Therefore, for $d < \sqrt{ \Delta_c-1 }$ the optimal choice amounts to the width of the support,
i.e.~the expression provided by the solid green line in \Fig{triangular-bounds}.

In the next section we extend our analysis to unimodal overshoots.

\subsection{ Bounds for unimodal overshoots }
\label{sec:bounds4unimodal}

\begin{figure}[t]
  \centering
  \includegraphics{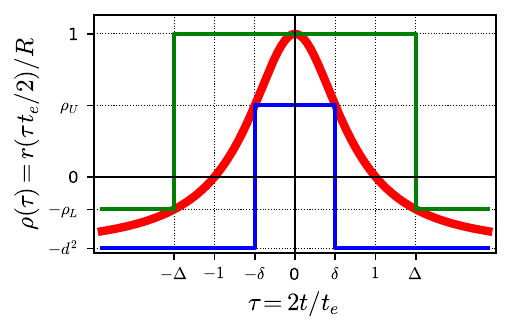}
  \caption{\label{fig:power-law-sketch}
    Notations adopted to derive optimal bounds for general unimodal overshoots.
  }
\end{figure}

In \Fig{TH_comparison} we announced
that the exponent of the decay for large $H$ is affected 
by a slow decay of $r(t)$ towards $-R_d$.
In order to describe this dependence we consider general overshoots of unimodal form
where $r(t)$ approaches $-R_d$ for $|t| \to \infty$,
and it takes a maximum value of $R$. 
To simplify notations we introduce the dimensionless time $\tau = 2 t/t_e$,
and we describe the overshoot by a function $R \, \rho(\tau)$
that provides the value of $r$ when the width of $r(t)$ amounts to $\tau\,t_e/2$.
For symmetric overshoots this amounts to
\begin{align*} 
  \rho(\tau) = R^{-1} \; r \left( \frac{\tau \, t_e}{2} \right) .
\end{align*}
The notations are summarized in \Fig{power-law-sketch}.

\subsubsection{Lower bound}

To derive a lower bound for unimodal overshoots we consider rectangular overshoots
that take a value of $R$ for $|\tau| < \tau_L$ with $\tau_L>1$,
and the value $R \, \rho(\tau_L)$ otherwise.
The green line in \Fig{power-law-sketch} provides the optimal case 
$\tau_L = \Delta$ and $\rho_L = -\rho(\Delta)$
that provides the maximum 
\begin{align}  \label{eq:cL_1step}
   c_L = \max_{\tau_L > 1} \left[   \frac{2}{\tau_L} \; \text{atan} \sqrt{ -\rho(\tau_L) } \right]
  \, .
\end{align}
The maximum is provided by the solution of 
\begin{align} \label{eq:general-cL_implicit}
  \sqrt{ \rho_L } \; \text{atan} \sqrt{ \rho_L }
  = - \frac{\Delta}{ 2 } \; \frac{\rho'(\Delta)}{ 1+\rho_L } \, .
\end{align}

For large $d$ we expect that the argument of the arcus tangent is large such that
\begin{subequations} \label{eq:general-cL}
\begin{align} \nonumber
  -\frac{\Delta}{2} \; \frac{\rho'(\Delta)}{1+\rho_L}
  &=
    \sqrt{\rho} \; \text{atan} \sqrt{\rho_L}
  \simeq
    \frac{\pi}{2} \; \sqrt{\rho_L} - 1
  \\[2mm]
  \label{eq:general-cL_large}
  \Rightarrow \quad
  \sqrt{\rho_L}
  &= \frac{2}{\pi} \; \left(
    1 -\frac{1}{2} \; \frac{\Delta \; \rho'(\Delta)}{1+\rho_L}
    \right)
    \quad \text{ for } \quad \rho_L \gg 1 \, .
\end{align}
The bound $c_L^>$ is obtained by evaluating this expression for the a given overshoot.

For small $d$ we expect that the argument of the arcus tangent in \Eq{general-cL_implicit} is small
such that
\begin{align}  \label{eq:general-cL_small}
  \rho_L = -\frac{1}{2} \; \Delta \; \rho'(\Delta) 
    \quad \text{ for } \quad \rho_L \ll 1 \, .
\end{align}
\end{subequations}
The bound $c_L^<$ is obtained by evaluating this expression for the a given overshoot.
Examples will be provided in \Secs{kappa} and \sect{Gaussian}.

\subsubsection{Upper bound}

To derive an upper bound we consider a rectangular overshoot
that takes a value of $r = R \, \rho(\tau_U)$ for $|\tau| < \tau_U < 1$ and the value $r = -R_d$ otherwise.
The blue line in \Fig{power-law-sketch} provides the optimal case
$\tau_U = \delta$ and $\rho_U = \rho(\delta)$
that provides the minimum 
\begin{align} \label{eq:general-cU}
  c_U = \min_{0 < \delta < 1} \left[ \frac{2}{\delta \: \sqrt{\rho_U}} \; \text{atan} \frac{d}{\sqrt{\rho_U}} \; \right]
  \quad \text{with } \rho_U = \rho(\delta) \, .
\end{align}

For large $d$ we expect that the argument of the arcus tangent is large.
Approximating it by its asymptotic value $\pi/2$ and taking the derivative with respect to $\delta$
provides the condition
\begin{subequations} \label{eq:general-cU_limits}
\begin{align} \label{eq:general-cU_large}
  \rho_U = -\frac{1}{2} \; \delta \; \rho'(\delta) 
    \quad \text{ for } \quad d \gg \sqrt\rho_U  \, .
\end{align}
The bound $c_U^>$ is obtained by evaluating this expression for the a given overshoot.

For small $d$ we expect that the argument of the arcus tangent is small.
Approximating it to linear order and taking the derivative with respect to $\delta$
provides the condition
\begin{align} \label{eq:general-cU_small}
  \rho_U = - \delta \; \rho'(\delta) 
    \quad \text{ for } \quad d \ll \sqrt\rho_U  \, .
\end{align}
\end{subequations}
The bound $c_U^<$ is obtained by evaluating this expression for the a given overshoot.

The shape of the boundaries will depend on the form of the overshoot, $\rho(\tau)$.
For the analysis it is useful to observe that \Eqs{general-cL_small} and \eqref{eq:general-cU_limits}
take the form
\begin{align}   \label{eq:general_implicit_condition}
  \rho(\tau) = - f \; \tau \; \rho'(\tau) 
\end{align}
with $f = -1/2$, $1/2$, and $1$, respectively.
Apart from the value of $f$ the cases only differ in the valid range of $\tau$,
and the limits to be taken when evaluating the arguments.

In the remainder of this section we work out the limits for Gaussian overshoots and for overshoots with power-law tails.

\section{ Nontrivial examples }
\label{sec:examples}

\subsection{ Gaussian Overshoots }
\label{sec:Gaussian}

Earlier literature\cite{Ritchie} addressed Gaussian overshoots.
In our notations they take the form
\begin{align} \label{eq:Gaussian-overshoots}
  \rho(\tau) = \frac{ r(t) }{R}
  = d^2 \;\; \left[ -1 + \left( \frac{1+d^2}{d^2} \right)^{1-\tau^2} \right] 
\end{align}
with derivative
\begin{align*} 
  \tau \: \rho'(\tau)
  = -2 d^2 \tau^2 \; \ln \left( \frac{1+d^2}{d^2}  \right) \;
  \left(  \frac{1+d^2}{d^2}  \right)^{1-\tau^2}
\end{align*}

\begin{figure}[t]
  \centering
  \includegraphics{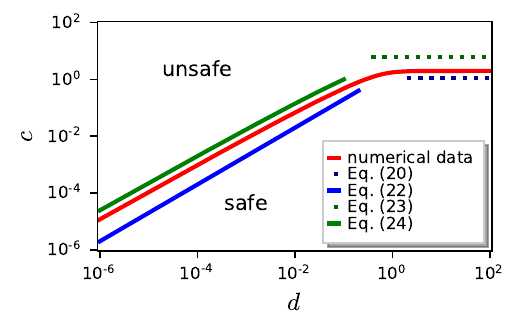}
  \caption{\label{fig:overshoot_Gauss_dc}
    The stability border for Gaussian overshoots, as defined in \Eq{Gaussian-overshoots}.
    The thick red line shows numerical data for the boundary.
    The other lines show upper and lower bounds for large and small values of $d$, respectively,
    as indicated in the legend.
  }
\end{figure}

For $d \gg 1$ and $d \gg \rho_L$ the expressions \eqref{eq:Gaussian-overshoots} imply
\begin{align*} 
  \rho_L &= d^2 \;\; \left[ 1 - \exp\left( \frac{1-\Delta^2}{d^2} \right) \right] = \Delta^2 - 1
  \\
  \Delta \: \rho'(\Delta)
  &= -2 \, \Delta^2
\end{align*}
such that the implicit equation \eqref{eq:general-cL_large} takes the form
\begin{align*} 
  \sqrt{\rho_L}
  = \frac{2}{\pi} \; \left(
    1 -\frac{1}{2} \; \frac{ \Delta \; \rho'(\Delta) }{1+\rho_L}
                               \right)
  = \frac{4}{\pi} 
  \, .
\end{align*}
Consequently, we obtain
\begin{align}   \label{eq:gauss_cL_large}
  c_L^> = \frac{2\: \arctan\sqrt{\rho_L}}{\Delta}
  = \frac{2\: \arctan\frac{4}{\pi}}{\sqrt{1 + \left( \frac{4}{\pi} \right)^2 }}
  \simeq 1.118
\end{align}
as opposed to the numerically obtained value of $2.01$.
Indeed, in \Fig{overshoot_Gauss_dc} there only is a small difference between the numerical data (solid red line)
and the prediction \Eq{gauss_cL_large}
that is provided by a dashed blue line.

We inspect \Eq{general_implicit_condition} to find the other bounds of the stability boundary.
For the distribution \Eq{Gaussian-overshoots} it takes the form
\begin{align*} 
  -1 & + \frac{1+d^2}{d^2} \;\exp\left[ -\tau^2 \; \ln\left( \frac{1+d^2}{d^2} \right) \right]
  \\[2mm]
  &= \frac{ \rho(\tau) }{d^2}
  = - f \; \frac{ \tau \; \rho'(\tau) }{d^2}
  \\[2mm]
  &= 2f \, \tau^2 \; \ln\left( \frac{1+d^2}{d^2} \right) \; \frac{1+d^2}{d^2} \; \exp\left[ -\tau^2 \; \ln\left( \frac{1+d^2}{d^2} \right) \right] .
\end{align*}
Introducing
\begin{subequations} \label{eq:Gauss_implicit_condition}
\begin{align}   \label{eq:Gauss_implicit_X}
  X = \tau^2 \; \ln\left( \frac{1+d^2}{d^2} \right) 
\end{align}
allows us to write this condition in the compact form
\begin{align}   \label{eq:Gauss_implicit_compact}
  \frac{d^2}{1+d^2} = ( 1 - 2\,f\: X ) \; \rme^{-X} \, .
\end{align}
\end{subequations}

In order to evaluate the small $d$ limit for the lower bound $c_L^<$,
we evaluate \Eq{Gauss_implicit_condition} for $f=-1/2$.
In this case the first factor of the right-hand-side of \Eq{Gauss_implicit_compact}
is larger than one, and the product can only be small when $X$ is large.
The logarithm of \Eq{Gauss_implicit_compact} provides
\begin{align*} 
  X
  &= \ln\left( \frac{1+d^2}{d^2} \right) + \ln( 1+X )
  \\[2mm]
  &\simeq \ln\left( \frac{1+d^2}{d^2} \right) + \ln\left[ 1+ \ln\left( \frac{1+d^2}{d^2} \right) \right] .
\end{align*}
Based on \Eq{Gauss_implicit_X} with $\tau = \Delta$ we find,
\begin{align*} 
  \Delta^2
  = \frac{X}{\ln\left( \frac{1+d^2}{d^2} \right) }
  = 1 + \frac{ \ln\left[ 1+ \ln\left( \frac{1+d^2}{d^2} \right) \right] }{ \ln\left( \frac{1+d^2}{d^2} \right) }
\end{align*}
and we obtain 
\begin{align*} 
  \rho_L
  & = -\rho(\Delta)
  = d^2 \; \left[ 1 - \exp \left( (1-\Delta^2) \; \ln\left( \frac{1+d^2}{d^2} \right) \right)  \right]
  \\[2mm]
  &= d^2 \; \left[ 1 - \frac{1}{ 1 + \ln\left( \frac{1+d^2}{d^2} \right) } \right]
    = \frac{d^2}{ 1 + \frac{1}{ \ln\left( \frac{1+d^2}{d^2} \right)} }
    \, .
\end{align*}
We see that $\Delta$ is larger than one, and $\rho_L$ is small.
Consequently,  $\text{atan}(\sqrt\rho_L) \simeq \sqrt\rho_L$ and we find
\begin{align}    \label{eq:gauss_cL_small}
  c_L^<
  &= \frac{2}{\Delta} \; \sqrt\rho_L
    \simeq 2 \: d
    \, .
\end{align}
This bound is shown by the solid blue line in \Fig{overshoot_Gauss_dc}.

The large $d$ limit for the upper bound is obtained by evaluating \Eq{Gauss_implicit_condition} for $f=1/2$,
and observing that $\ln( 1 + d^{-2} ) \simeq d^{-2}$.
Therefore, $X \simeq (\delta/d)^2 \ll 1$ since $0 < \delta < 1$,
Hence, \Eq{Gauss_implicit_compact} reduces to
\begin{align*} 
  1 - \frac{1}{d^2} = \frac{d^2}{1+d^2} = ( 1 - X ) \; \rme^{-X}
  = 1 - 2\, \frac{\delta^2}{d^2} 
\end{align*}
such that
\begin{align*} 
  \delta^2
      \simeq \frac{1}{2}
  \quad\text{and}\quad
  \rho_U
  = d^2 \; \left[ -1 + \exp \left( \frac{1-\delta^2}{d^2} \right) \right]
  = \frac{1}{2} \, .
\end{align*}
Consequently, we find
\begin{align}  \label{eq:gauss_cU_large}
  c_U^> = \frac{2}{\delta \, \sqrt\rho_U} \; \text{atan} \frac{d}{\sqrt\rho_U}
  = 2\pi \, .
\end{align}
This bound is shown by the dotted green line in \Fig{overshoot_Gauss_dc}.

Finally, the small  $d$ limit for the upper bound is obtained by evaluating \Eq{Gauss_implicit_condition} for $f=1$.
In this case $X$ can not exceed the value of $1/2$ because the expression on the right hand side of \Eq{Gauss_implicit_compact} must be positive.
Hence, the exponential can not be small such that
\begin{align*} 
  X = \frac{1}{2} - d^2 \: \sqrt{e} + \mathcal{O}(d^2) \, .
\end{align*}
In that case 
\begin{align*} 
  \left( \frac{1+d^2}{d^2} \right)^{-\delta^2} &= \rme^{-X} = \frac{1}{\sqrt\rme}
  \\
  \text{and}\quad
  \delta^2 &= \frac{1}{2 \; \ln\left( \frac{1+d^2}{d^2} \right)}
  \simeq \frac{1}{ 4 \, \left\lvert \ln d \right\rvert }
\end{align*}
such that
\begin{align*} 
  \rho_u
  = d^2 \: \left( -1 + \frac{1+d^2}{d^2} \frac{1}{\sqrt\rme} \right)
  \simeq \frac{1}{\sqrt\rme}
  \, .
\end{align*}
Consequently, we obtain
\begin{align}   \label{eq:gauss_cU_small}
  c_U^<
  = \frac{2 \: d}{\delta \: \rho_U}
  = 4 \, \sqrt{e} \;\; d \: \sqrt{ \left\lvert \ln d \right\rvert }
  \, .
\end{align}
This bound is shown by the solid green line in \Fig{overshoot_Gauss_dc}.
One clearly sees that the numerical data follow the $d \, \sqrt{|\ln d|}$
dependence provided by \Eq{gauss_cU_small},
rather than the linear dependence of \Eq{gauss_cL_small}.
This suggests that the decay of the tails of the unimodal distribution
might have an impact on the crossover from constant for large $d$
to decaying for decreasing small $d$ behavior.
We will further explore that possibility by inspecting overshoots
with power-law tails.

\subsection{ Overshoots with power-law tails }
\label{sec:kappa}

As an examples of an overshoot with very broad, power-law tails we inspect overshoots of the form
\begin{subequations} \label{eq:power-law-overshoots}
\begin{align} 
  \rho(\tau) 
  &= d^2 \;\; \frac{ 1 - \left\vert \tau \right\rvert^{\kappa} }{ d^2 + \left\vert \tau \right\rvert^{\kappa} }  
\end{align}
with derivative
\begin{align}  \label{eq:power-law-overshoot_derivative}
  \tau \: \rho'(\tau)
  = - d^2 \; \frac{1+d^2}{ \left( \left\vert \tau \right\rvert^\kappa + d^2 \right)^2 } \; \kappa \; \left\vert \tau \right\rvert^\kappa
  \, .
\end{align}
\end{subequations}
For $\kappa =2$ this amounts to a Cauchy distribution with base line $-d^2$ (thick red line \Fig{power-law-sketch}).
The stability borders for $\kappa \in \{ 2,3,4 \}$ are plotted in \Fig{TH_comparison}.

\begin{figure*}[t]
  \[
    \includegraphics{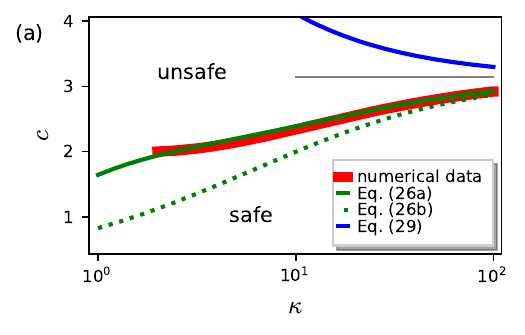} \qquad
    \includegraphics{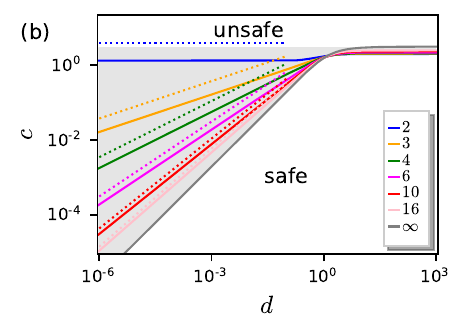}
  \]
  \caption{ \label{fig:algebraic_borders}
    Comparison of predictions and numerical data for the stability border for overshoots with power-law tail, 
    as defined in \Eq{power-law-overshoots}.
    \\
    (a) The $\kappa$ dependence of the position $c(d)$ of the stability boundary for $d=100$.
    The red line provides numerical data for $\kappa > 2$.
    It amounts to the stability boundary between the safe and the unsafe region.
    The green lines are (approximations to the) lower bound $c_L^>$ discussed in the main text,
    the thin gray line provides the asymptotic value $\lim_{\kappa\to\infty} c(d) = \pi$,
    and the blue line is the upper bound $c_U^>$, \Eq{kappa_-cU_large}.
    \\
    (b) The thin solid lines show numerical results for the boundaries $c(d)$ with
    different $\kappa$, as provided in the figure legend, i.e.~for
    $\kappa \in \{2,3,4,6,10,16, \infty\}$ from top to bottom.
    The corresponding upper bounds for the decay for small $d$, \Eq{kappa_-cU_small},
    are provided by dotted lines with matching colors.
    The $\kappa$ dependence of the asymptotic value to the right is shown in panel (a),
  }
\end{figure*}

For $d \gg 1$ the implicit equation \eqref{eq:general-cL_large} takes the form
\begin{align*} 
  \sqrt{ \Delta^\kappa - 1 } &=
  \sqrt{\rho_L}
  = \frac{2}{\pi} \; \left(
    1 -\frac{1}{2} \; \frac{ \Delta \; \rho'(\Delta) }{1+\rho_L}
                               \right)
  = \frac{2}{\pi} \; \left( 1 + \frac{\kappa}{2} \right) .
\end{align*}
This provides $\Delta$, and hence the bound
\begin{subequations}
\begin{align} \label{eq:kappa_-cL_largeE}
  c_L^>
  &= \frac{\pi}{\Delta}
  = \frac{\pi}{ \left( 1 + \frac{4}{\pi^2} \; \left( 1 + \frac{\kappa}{2} \right)^2 \right)^{1/\kappa} }
  \, .
\end{align}
Note that we dropped terms of order $\rho_L^{-1}$ in the derivation of \Eq{general-cL_large}.
Hence, this expression holds at best when $\kappa$ is not too small.
Indeed, numerical data reveal that this approximation increases the prediction \Eq{kappa_-cL_large}
by about $70$\% for $\kappa = 2$ and
still $20$\% for $\kappa = 10$. 
As a consequence, $c_L^>$ is not even a lower bound,
but rather it is larger than the boundary for all $\kappa \geq 3$.
However, it never exceeds the boundary by more than $2$\%,
and the relative deviation decays like $\kappa^{-1}$.
The solid green line in \Fig{algebraic_borders}(a) demonstrates
that it provides an excellent estimate of the asymptotic value of the border for large $d$.
A proper lower bound can be found by numerically solving the large $d$ approximation of the implicit equation \eqref{eq:general-cL_large}
\begin{align}  \label{eq:kappa_-cL_large}
  \sqrt\rho_L \; \text{atan}\sqrt\rho_L = \frac{\kappa}{2} \, .
\end{align}
The resulting expression is shown by the dotted line in \Fig{algebraic_borders}(a).
\end{subequations}

We inspect \Eq{general_implicit_condition} to find the other bounds of the stability boundary.
For the distribution \Eq{power-law-overshoots} it takes the form
\begin{align*} 
  \frac{ 1 - \left\vert \tau \right\rvert^{\kappa} }{ d^2 + \left\vert \tau \right\rvert^{\kappa} }  
  = \frac{ \rho(\tau) }{d^2}
  = - \frac{ f \; \tau \; \rho'(\tau) }{ d^2 }
  = \frac{1+d^2}{ \left( \tau^\kappa + d^2 \right)^2 } \; \kappa \: f \; \tau^\kappa \, .
\end{align*}
This provides a quadratic equation for $\tau^\kappa$ with solutions
\begin{align}  \nonumber
  \tau^\kappa_\pm
  &= \frac{1}{2} \; \left[ \left( 1-\kappa\:f \right) - \left( 1+\kappa\:f \right) \; d^2  \right] 
  \\[2mm] \label{eq:algebraic_tau_pm}
  &\quad \times  \left[ 1 \pm \sqrt{ 1 + \frac{4 \, d^2}{ \left[ \left( 1-\kappa\:f \right) - \left( 1+\kappa\:f \right) \; d^2  \right]^2 }} \right] .
\end{align}

In order to evaluate the small $d$ limit for the lower bound $c_L^<$,
we evaluate \Eq{algebraic_tau_pm} for $f=-1/2$.
For small $d$ the solution $\tau^\kappa_-$ has a negative sign, and it must be discarded.
Consequently,
\begin{align*} 
  \Delta^{\kappa} = 1 + \frac{\kappa}{2}
  \qquad\text{and}\qquad
  \rho_L
  \simeq d^2 \; \frac{\Delta^\kappa - 1}{\Delta^\kappa}
  = d^2 \frac{ \frac{\kappa}{2} }{1 + \frac{\kappa}{2}}
\end{align*}
and we find
\begin{align} \label{eq:kappa_-cL_small}
  c_L^<
  &= \frac{2}{\Delta} \; \sqrt{ \rho_L }
  = 2 \; d \; \sqrt{\frac{\kappa}{2}} \;\; \left( 1 + \frac{\kappa}{2} \right)^{-\frac{1}{2}-\frac{1}{\kappa}}
  \, .
\end{align}
These lower bounds are linear functions that provide only a very rough lower bound.
However this bound can be improved noticeably by introducing another step in the piecewise-constant function adopted to derive the bound (see \Fig{algebraic_2step-lower-borders}).
We provide the according derivation in \App{2step}.

For large $d$ the solution $\tau^\kappa_+$ has a negative sign, and it must be discarded.
Moreover, the term added to one under the square root is now of order $d^{-2}$.
Expanding the square root one obtains
\begin{align*} 
  \delta^{\kappa} = \frac{1}{1 + \frac{\kappa}{2}}
  \qquad\text{and}\qquad
  \rho_U
  \simeq 1 - \delta^\kappa
  = \frac{ \frac{\kappa}{2} }{1 + \frac{\kappa}{2}} \, .
\end{align*}
In view of \Eq{general-cU} we find
\begin{align} \label{eq:kappa_-cU_large}
  c_U^>
  &= \frac{\pi}{ \delta \; \sqrt{ \rho_U } }
    =
    \frac{ \pi }{\sqrt{\kappa/2}} \; \left( 1 + \frac{\kappa}{2} \right)^{\frac{1}{2}+\frac{1}{\kappa}} 
  \, .
\end{align}
It is shows as solid blue line in \Fig{algebraic_borders}(a).

Finally, the small  $d$ limit for the upper bound is obtained by evaluating \Eq{algebraic_tau_pm} for $f=1$.
For $\kappa > 1$ the solution $\tau^\kappa_+$ has a negative sign, and it must be discarded.
Moreover, the term added to one under the square root is now of order $d^2$.
Expanding the square root one obtains
\begin{align*} 
  \delta^{\kappa} = \frac{d^2}{\kappa - 1}
  \qquad\text{and}\qquad
  \rho_U
  \simeq \frac{1}{1 + \frac{1}{\kappa-1}}
  = \frac{\kappa-1}{\kappa}
  \, ,
\end{align*}
and this provides
\begin{align} \label{eq:kappa_-cU_small}
  c_U^<
  &= \frac{2 \; d}{ \delta \; \rho_U }
    =
    \frac{\kappa}{( \kappa - 1 )^{1-\frac{1}{\kappa}}} \;2\; d^{1-\frac{2}{\kappa}}
  \, .
\end{align}
These upper bounds are power laws in $d$ with exponents $1-2/\kappa$.
\Fig{algebraic_borders}(b) shows that this dependence is also observed in the numerical data.
Moreover, for moderately large $d$ the predicted prefactors approach those of the numerical data
such that \Eq{kappa_-cU_small} provides an accurate estimate of the stability border.

In summary, we find that for large values of $d$ the boundaries lies close to $\pi$.
This behavior was also reported in earlier work.\cite{Ritchie}
However, for small values of $d$ the upper bound $c_U^<$ implies
that the boundaries follow power laws with exponents $1-2/\kappa$.
For $\kappa=2$ there is no change.
However, for $\kappa > 2$ the unsafe region is substantially increasing in this case,
as shown in \Fig{algebraic_borders}(b).

\section{Discussion}
\label{sec:discussion}

In the present section we revisit results reported in earlier literature,
and we discuss the impact of our findings for applications.

\subsection{Earlier literature}

\citeauthor{Ritchie}\cite{Ritchie} reported the following condition for the safe region
\begin{align}
  d^b\,R\, t_e^2 \leq 16
\label{eq:ritchie}
\end{align}
where $d^b$ is the ratio of the square of decay rate $\lambda$ to the stable fixed point and the position of the fixed point,
$-x_c$, evaluated right at the bifurcation value,
\begin{align*} 
  \lambda
  = -\left . \frac{\rmd \dot x}{\rmd x} \right\rvert_{x=-x_c} \!\!
  = 2k\,x_c = 2 \, \sqrt{k\,|r|}
  \quad\Rightarrow\quad
  d^b = \lim_{r \nearrow 0} \frac{\lambda^2}{|r|} = 4\, k \, .
\end{align*}
Hence, \Eq{ritchie} is equivalent to
\begin{align*} 
  T
  = \sqrt{ k R_d \, t_e^2 }
  \leq \sqrt{ k R_d } \: \frac{4}{\sqrt{ d^b \: R}}
  = \frac{2}{\sqrt H}
\end{align*}
as provided by the uppermost, black dotted line in \Fig{TH_comparison}.
Moreover, this also amounts to $c = T \, \sqrt{H} \leq 2$
which amounts to the horizontal, large $d$ behavior observed in
\Figs{triangular-bounds},~\fig{overshoot_Gauss_dc}, and~\fig{algebraic_borders}.

The relation to our present work becomes transparent when adopting the dimensionless coordinates
$\hat x = x / \sqrt{R_d/k}$ and $\tau = t / (T \, t_e)$
such that \Eq{EOM} takes the form
\begin{align*} 
  \frac{\rmd \hat x}{\rmd \tau} = \hat x^2 + H \: \frac{r(t)}{R} \, .
\end{align*}
The derivation of \Eq{ritchie} assumes
that the forcing term is approximately parabolic in $[-t_e/2, t_e/2]$, and
that the parameter forcing is of the form $\epsilon \, R_0 + r_h(\epsilon\, t)$
[cf.~Eq.~(2.2) and the discussion below Eq.~(2.10) in \citeauthor{Ritchie}\cite{Ritchie}],
where $R_0$ and $r_h(\epsilon\, t)$ take values of order one.
The former condition implies that
\begin{align*} 
  H \: \frac{r(t)}{R}
  = H \: \left( 1 - \left( \frac{2t}{t_e} \right)^2 \right)
  = H - 4 \; \left( \sqrt{H} \; \frac{\tau}{T} \right)^2 \, ,
\end{align*}
and the latter condition stipulates that both $H$ and $\sqrt{H} \, T$ take values of order $\epsilon \ll 1$.
Hence,
\begin{align*} 
  T
  \lesssim \frac{\sqrt{H}}{\epsilon} \sim \frac{1}{\sqrt H}
\end{align*}
is an immediate consequence of the adopted scaling.

The condition, \Eq{ritchie}, was derived based on the condition
that $R$ is small and $t_e$ is large.
It does not involve the asymptotic values $R_d$.
Here we establish that the adopted scaling holds when $H \lesssim 1$,
i.e.~for $R \lesssim R_d$.
Moreover, \citeauthor{Ritchie}\cite{Ritchie} assume
that the control parameter takes a parabolic form beyond the threshold.
The bounds provided in \Sect{bounds} establish
that only minor changes of the numerical factor emerge in \Eq{ritchie}
when one drops this requirement.

\subsection{Applications in Earth system sciences}
\label{sec:climate-application}

\FIG{TH_comparison} entails that critical threshold $H_c$ for short transgressions of duration $T$ over the tipping point
is substantially smaller than previously expected:
Previous work\cite{Ritchie} suggested that $H_c \simeq 4/T^{2}$ 
while  \Eq{c-jump-boundary} entails a cross-over to linear scaling $c \sim d$ for small $c$ and $d$.
This implies 
\begin{align}   \label{eq:critH}
  H_c \sim \frac{1}{T} \ll \frac{4}{T^2} \, .
\end{align}
We provide two examples that indicated how this might be important in Earth system sciences.

The Earth climate system comprises subsystems that may exhibit threshold behavior.\cite{drouet2021net,riahi2021cost,rogelj2019new}
These are the so-called climate tipping elements,\cite{tippingbasics,armstrong2022exceeding}
which can conceptually be represented through models that exhibit a saddle-node bifurcation (tipping point)
such as the Amazon rainforest, the large ice sheets on Greenland and Antarctica, or the Atlantic Meridional Overturning Circulation.\cite{staal2015synergistic,levermann2016simple,stommel1961thermohaline,cessi1994simple,wunderling2021modelling,2021KohlerWunderlingDongesVollmer}
The risk of at least temporarily transgressing such critical thresholds of the Earth system increases as global warming levels increase.\cite{Ritchie2,wunderling2023global}.
Indeed, it is unlikely
that the Paris climate targets of limiting global warming to, at best, \SI{1.5}{\celsius} by the end of this century can be satisfied without an overshoot.\cite{masson2018global,schleussner2022emission,rogelj2023credibility-gap} 
This will put some climate tipping elements at risk of losing their stability,\cite{wunderling2023global} 
and may ultimately lead to severe biosphere degradation or large levels rises of sea level on the timescales of centuries and beyond.
As compared to typical relaxation time scales of the climate system such an overshoot will be short,
and according to \Eq{critH} the maximum admissible trespassing $H_c$ is substantially smaller than expected.

Shallow landslides and debris flow are triggered by intense and/or long-lasting rainfalls.\cite{1980Caine,1993LarsenSimon,2017ChaeParkCataniSimoniBerti}
The stability threshold of slope stability has been described as a power law $I \sim D^{-\gamma}$
where $I$ is the intensity of the rainfall and $D$ the duration;
with exponents $\gamma$ in the range between $0.4$ (relatively undisturbed slopes at many different sites\cite{1980Caine}) and $0.8$ (humid-tropical environment in Puerto Rico\cite{1993LarsenSimon}).
A comprehensive recent overview of data is provided by Fig.~7 of Ref.~\citenum{2017ChaeParkCataniSimoniBerti}. 
The present study suggests that it might be necessary to describe this data by a cross-over between two power laws.
Further work is needed to identify the appropriate nondimensionalized parameters.
We expect that future data availability will enable better assessments of the risks of slope instabilities,
in particular when the frequency of extreme rainfall events increases with climate change (cf.~Fig.~6 of Ref.~\citenum{masson2021ipcc}).

\section{Conclusion}
\label{sec:conclusion}

In this article we discussed the boundary separating safe and unsafe overshoots
when crossing a saddle-node bifurcation.
For a piecewise-constant evolution of the control parameter we derived an analytical solution for the boundary,
\Eq{jump-boundary}.
Based on a monotonicity condition, \Eq{monotonicity-condition},
this result can be used to obtain rigorous upper and lower bounds of the boundary for different profiles of the control parameter.
The conditions are formulated in terms of the dimensionless strength, $H$, and duration, $T$, of the overshoot --- 
or equivalently in terms of $c = T \, \sqrt H$ and $d = 1/\sqrt H$
which turns out to be more convenient for the calculations.

For small $H$ \citeauthor{Ritchie}\cite{Ritchie} established that the region of safe overshoots is bounded by a power law,
\begin{align*} 
  T < \frac{2}{ \sqrt H }
  \qquad\text{for} \quad
  H \lesssim 1 \, .
\end{align*}
Here, we extended this result by the following findings:

\citeauthor{Ritchie}\cite{Ritchie} assume that the control parameter takes a parabolic form beyond the threshold.
We establish that, up to minor changes of the prefactor, the power law applies for $H \lesssim 1$,
irrespective of the functional form of the time dependence of the order parameter. 
Moreover, for $H \gtrsim 1$ there is a crossover to a different power law
with an exponent that depends on the asymptotic decay of the control-parameter profile towards $R_d$.

For control-parameter profiles with a finite support we find that the safe region is bounded by a power law
that decays much more rapidly,
\begin{align*} 
  T < \frac{c}{\sqrt H} \simeq \frac{2\phi }{H}
  \quad \text{for} \quad
  H \gtrsim 1 \, ,
  \text{ finite support} 
\end{align*}
where $\phi$ is a constant that depends on the shape of the profile (cf.~\Eqs{bound-triangle}, \eqref{eq:boundary-cd}, and \Fig{triangular-bounds}).
It takes values slightly larger than one.

For a Gaussian shape of the control parameter the safe region is bounded by (\cf~\Eq{gauss_cU_small})
\begin{align*} 
  T 
  \lesssim \frac{4\,\sqrt\rme \: \sqrt{|\ln d|} }{H}
  = 4\: \sqrt{2\rme} \; \frac{\sqrt{\ln H}}{H}
  \quad \text{for} \quad
  H \gtrsim 1 \, ,
  \text{ Gaussian}
\end{align*}
as shown in \Figs{TH_comparison} and \fig{overshoot_Gauss_dc}.

For a large-$t$ asymptotics of $r(t)-R_d \sim t^{-\kappa}$ \Eqs{boundary-cd} and \eqref{eq:kappa_-cU_small} imply
that
\begin{align*} 
  T 
  \lesssim  \frac{d^{1-\frac{2}{\kappa}}}{\sqrt H} = H^{-1+\frac{1}{\kappa}}
  \quad \text{for} \quad
  H \gtrsim 1 \, ,
  \text{ power law}
\end{align*}
as shown in \Figs{TH_comparison}, \fig{algebraic_borders}, and \fig{algebraic_2step-lower-borders}. 

Finally, our approach is not limited to control-parameter profiles, $r(t)$,
that are symmetric with respect to the maximum.
In \Sect{bounds4unimodal} we explain how the bounds are constructed for arbitrary unimodal, i.e.~also asymmetric, profiles
(see also \Figs{triangular-bounds}).

We conclude that overshoots are less safe than previously thought.
Especially, high and fast overshoots are more likely to trigger a tipping event (see \Fig{TH_comparison}).
This emendation of the  earlier literature is of pivotal importance
when examining tipping risks in applications of nonlinear dynamics in finance, ecology or climate.\cite{brummitt2015coupled}
For instance, in the field of climate tipping elements, assessing whether certain global warming overshoot trajectories can be considered safe or unsafe and which tipping risks are associated with them,
is key for avoiding potentially dangerous climate change pathways \cite{kemp2022climate} (cf.~\Sect{climate-application}).
The present results provide the foundation to establish rigorous boundaries for the risks in different model scenarios.

\begin{acknowledgments}
  We are thankful for the fruitful discussions with Paul Ritchie, Jonathan Donges and Vitus Benson.
  N.W. acknowledges support from the European Research Council Advanced Grant project ERA (Earth Resilience in the Anthropocene, ERC-2016-ADG-743080).
  E.E. is grateful for the support of the German Academic Scholarship Foundation and the funding through the postgraduate scholarship granted by the Free State of Saxony.
\end{acknowledgments}

\appendix
\section{Improved lower bound for overshoots with power-law tails}
\label{app:2step}

\begin{figure*}[t]
  \[
    \includegraphics{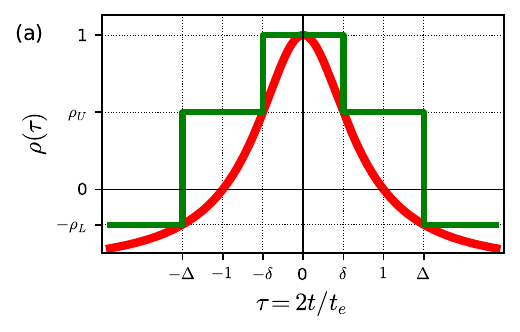} \qquad
    \includegraphics{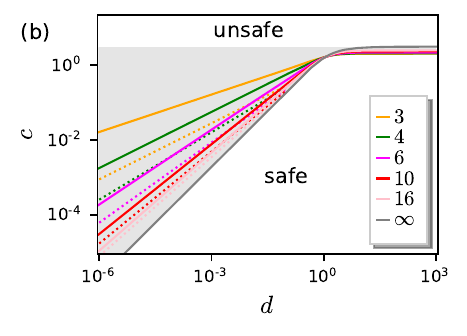}
  \]
  \caption{ \label{fig:algebraic_2step-lower-borders}
    Lower bound based on a two-step approximation for unimodal overshoots.
    (a) Notations adopted to parameterize the piecewise constant function evaluated to derive the improved lower bound.
    (b) Analogous plot to \Fig{algebraic_borders}(b),
    but now with the lower bounds provided by \Eq{kappa_cL_small_2step}.
    For large $\kappa$ the bound approaches $2d$, as we also observed earlier.
  }
\end{figure*}

\EQ{kappa_-cL_small} provides a lower bound to the stability border, $c(d)$,
that is a linear function of $d$,
while the numerical data indicate that it scales like $d^{1-2/\kappa}$.
We will demonstrate here how the bound can be improved by introducing another step in the piecewise-constant function adopted to derive the bound (see the green line in \Fig{algebraic_2step-lower-borders}(a) as opposed to the one in \Fig{power-law-sketch}).

In this case \Eq{solutiontan} still holds for $-\Delta < \tau \leq -\delta$,
and we use the resulting value of $x(-\delta t_e/2)$ as initial condition for further integration.
This provides
\begin{widetext}
\begin{align*} 
  \frac{x(\tau \, t_e/2)}{x_c} = \sqrt{H} \; \tan \left( \frac{1}{2} \, T \, \sqrt H \; (\tau + \delta)
  + \text{arctan}\left[ \sqrt\rho_U \; \tan \left(  T \sqrt{\rho_U H} \;\; \frac{\Delta - \delta}{2} - \text{arctan}\sqrt\frac{\rho_L}{\rho_U} \right) \right]
  \right)
  \qquad \text{for \ } -\delta \leq \tau < \delta \, ,
\end{align*}
\end{widetext}
where $\rho_U = \rho(\delta)$ and $\rho_L = -\rho(\Delta)$.

The stability boundary is provided by the trajectory that will approach the unstable fixed point for $\tau \to \infty$.
By time-reversal symmetry this trajectory proceeds through the origin such that $x(0) = 0$.
Hence, we obtain
\begin{align*} 
  \tan \frac{\delta \: T \sqrt H }{2}
  = -\sqrt\rho_U \; \tan \left( T \: \sqrt{\rho_U H} \;\; \frac{\Delta-\delta}{2}
  - \text{arctan} \sqrt{\frac{\rho_L}{\rho_U}} \right) .
\end{align*}
Introducing $c = T\sqrt H$ and $d=1/\sqrt H$ provides the lower bound $c_L$
as the smallest value of $c$ that is a solution of the implicit equation
\begin{align*} 
  \tan \frac{c\:\delta}{2}
  =
  -\sqrt{\rho_U} \; \tan \left( c\: \sqrt{\rho_U} \;\; \frac{\Delta-\delta}{2} - \text{arctan} \sqrt{\frac{\rho_L}{\rho_U}}  \right) .
\end{align*}
Here the minimization is performed by looking for the optimal choice of $\delta$ and $\Delta$.
The corresponding result for a single step was \Eq{cL_1step}.

We will now evaluate this implicit equation for small values of $c$,
and $\rho(\tau)$ provided in \Eq{power-law-overshoots}.
By definition we have $0 < \delta < 1$ such that the argument of the tangent of the left-hand side of the equation is small,
$c\, \delta/2 \ll 1$. 
To evaluate the right-hand side we observe that $\tan(a-b) = (\tan a - \tan b) / (1 + \tan a \; \tan b)$,
and that also $c\: \sqrt{\rho_U} (\Delta-\delta) \ll 2$ will be small.
This provides
\begin{align} \nonumber
  \frac{c\: \delta}{2 \: \sqrt{\rho_U}}
  &= \frac{ \sqrt \frac{\rho_L}{\rho_U} - c\: \sqrt{\rho_U} \;\; \frac{\Delta-\delta}{2} }
  { 1 + \sqrt \frac{\rho_L}{\rho_U} \;\; c\: \sqrt{\rho_U} \;\; \frac{\Delta-\delta}{2} }
  \\[2mm] \nonumber
  \Rightarrow\quad
  0
  &= c^2 \: \rho_L
    + 2 \, c \, \sqrt{\rho_L} \; \left( \frac{\rho_U}{\delta} + \frac{1}{\Delta-\delta} \right)
    - \frac{4 \: \rho_L}{\delta \: (\Delta-\delta)}
  \\[2mm]  \label{eq:cL_2step}
  \Rightarrow\quad
  c_L \:
  &\simeq \: \max_{\delta,\Delta}
    \frac{2 \: \sqrt{-\rho(\Delta)}}{ \delta + \rho(\delta) \: (\Delta-\delta) } \, .
\end{align}

The maximum will be taken for the value of $\delta$ the minimizes the denominator,
\begin{align*} 
  0 &= \frac{\partial}{\partial \delta} \Bigl[ \delta + \rho(\delta) \: (\Delta-\delta)   \Bigr]
  = 1 - \rho(\delta) + \left( \frac{\Delta}{\delta} - 1 \right) \; \delta \: \rho'(\delta) .
\end{align*}
We insert \Eq{power-law-overshoot_derivative}, observe that $\Delta \gg \delta$, and keep only leading-order terms in $d$:
\begin{align*} 
  \delta^{\kappa+1} = \Delta \: \kappa\: d^2
  \qquad \text{and} \qquad
  \rho_U = \rho(\delta) = \frac{\delta}{\Delta \: \kappa} \, .
\end{align*}

Next we evaluate the $\Delta$-derivative of \Eq{cL_2step},
\begin{align*} 
  0 &= \frac{\partial}{\partial \Delta} \frac{2 \: \sqrt{-\rho(\Delta)}}{ \delta + \rho(\delta) \: (\Delta-\delta) }
  \\[2mm]
  \Rightarrow\qquad
  2 \: \rho(\Delta) \: \rho_U
    &= \rho'(\Delta) \; \left( \delta \: (1-\rho_U) + \Delta \: \rho_U \right) .
\end{align*}
In view of $d^2 \ll 1 < \Delta$ this entails
\begin{align*} 
  \frac{\kappa}{\Delta^\kappa - 1}
  \simeq \frac{ \Delta \: \rho'(\Delta) }{ \rho(\Delta) }
  &= \frac{ 2 \Delta \: \rho_U}{ \delta \: (1-\rho_U) + \Delta \: \rho_U }
  \\[2mm]
  &= \frac{2}{\kappa + 1 - \frac{\delta}{\Delta}}
    \simeq \frac{2}{\kappa + 1}
\end{align*}
such that
\begin{align*} 
  \Delta^\kappa &= 1 + \frac{\kappa \: (1+\kappa)}{2}
                  \\[2mm]
  \text{and} \qquad
  \rho_L &= d^2 \; \frac{\Delta^\kappa - 1}{\Delta^\kappa}
  = d^2 \; \frac{\kappa \: (1+\kappa)}{2+\kappa\:(1+\kappa)} \, .
\end{align*}

Inserting these expressions into $c_L$ provides
\begin{align} 
  c_L^< \nonumber
  &\simeq \frac{2 \: \sqrt{\rho_L} }{ \delta + \Delta \: \rho_U }
  \\[2mm] \label{eq:kappa_cL_small_2step}
  &= d^{1 - \frac{2}{1+\kappa}} \;\; \sqrt{\frac{2\, \kappa}{1+\kappa}} \;\;
    \kappa^{1-\frac{1}{1+\kappa}} \;\; 
    \left( 1+ \frac{\kappa \: (1+\kappa)}{2} \right)^{-\frac{1}{2} - \frac{1}{\kappa\:(1+\kappa)}} \, .
\end{align}
In \Fig{algebraic_2step-lower-borders}(b) these bounds are compared to the numerical results.
The present two-step approach improves the lower bound from a linear function for all values of $\kappa$,
\Eq{kappa_-cL_small} to become a power-law in $d$ with an exponent $1 - 2/(1+\kappa)$.
This exponent is much closer to $1 - 2/\kappa$,
as observed in the numerical data and the upper bound, \Eq{kappa_-cU_small},
and --- in principle --- it can further be improved by introducing still another step in the piecewise constant function adopted to calculate the bound.

\section*{References}
\bibliography{saddlenode_bibliography,tipping}

\begin{thebibliography}{36}%
\makeatletter
\providecommand \@ifxundefined [1]{%
 \@ifx{#1\undefined}
}%
\providecommand \@ifnum [1]{%
 \ifnum #1\expandafter \@firstoftwo
 \else \expandafter \@secondoftwo
 \fi
}%
\providecommand \@ifx [1]{%
 \ifx #1\expandafter \@firstoftwo
 \else \expandafter \@secondoftwo
 \fi
}%
\providecommand \natexlab [1]{#1}%
\providecommand \enquote  [1]{``#1''}%
\providecommand \bibnamefont  [1]{#1}%
\providecommand \bibfnamefont [1]{#1}%
\providecommand \citenamefont [1]{#1}%
\providecommand \href@noop [0]{\@secondoftwo}%
\providecommand \href [0]{\begingroup \@sanitize@url \@href}%
\providecommand \@href[1]{\@@startlink{#1}\@@href}%
\providecommand \@@href[1]{\endgroup#1\@@endlink}%
\providecommand \@sanitize@url [0]{\catcode `\\12\catcode `\$12\catcode `\&12\catcode `\#12\catcode `\^12\catcode `\_12\catcode `\%12\relax}%
\providecommand \@@startlink[1]{}%
\providecommand \@@endlink[0]{}%
\providecommand \url  [0]{\begingroup\@sanitize@url \@url }%
\providecommand \@url [1]{\endgroup\@href {#1}{\urlprefix }}%
\providecommand \urlprefix  [0]{URL }%
\providecommand \Eprint [0]{\href }%
\providecommand \doibase [0]{http://dx.doi.org/}%
\providecommand \selectlanguage [0]{\@gobble}%
\providecommand \bibinfo  [0]{\@secondoftwo}%
\providecommand \bibfield  [0]{\@secondoftwo}%
\providecommand \translation [1]{[#1]}%
\providecommand \BibitemOpen [0]{}%
\providecommand \bibitemStop [0]{}%
\providecommand \bibitemNoStop [0]{.\EOS\space}%
\providecommand \EOS [0]{\spacefactor3000\relax}%
\providecommand \BibitemShut  [1]{\csname bibitem#1\endcsname}%
\let\auto@bib@innerbib\@empty
\bibitem [{\citenamefont {Horsthemke~W}(1984)}]{1984HorsthemkeLefever}%
  \BibitemOpen
  \bibfield  {author} {\bibinfo {author} {\bibfnamefont {L.~R.}\ \bibnamefont {Horsthemke~W}},\ }\href@noop {} {\emph {\bibinfo {title} {Noise-Induced Transitions: Theory and Applications in Physics, Chemistry, and Biology}}}\ (\bibinfo  {publisher} {Springer-Verlag},\ \bibinfo {address} {New York},\ \bibinfo {year} {1984})\BibitemShut {NoStop}%
\bibitem [{\citenamefont {Scheffer}\ \emph {et~al.}(2009)\citenamefont {Scheffer}, \citenamefont {Bacompte}, \citenamefont {Brock}, \citenamefont {Brovkin}, \citenamefont {Carpenter}, \citenamefont {Dakos}, \citenamefont {Held}, \citenamefont {van Nes}, \citenamefont {Rietkerk},\ and\ \citenamefont {Sugihara}}]{2009SchefferBacompteBrock-EtAl}%
  \BibitemOpen
  \bibfield  {author} {\bibinfo {author} {\bibfnamefont {M.}~\bibnamefont {Scheffer}}, \bibinfo {author} {\bibfnamefont {J.}~\bibnamefont {Bacompte}}, \bibinfo {author} {\bibfnamefont {W.~A.}\ \bibnamefont {Brock}}, \bibinfo {author} {\bibfnamefont {V.}~\bibnamefont {Brovkin}}, \bibinfo {author} {\bibfnamefont {S.~R.}\ \bibnamefont {Carpenter}}, \bibinfo {author} {\bibfnamefont {V.}~\bibnamefont {Dakos}}, \bibinfo {author} {\bibfnamefont {H.}~\bibnamefont {Held}}, \bibinfo {author} {\bibfnamefont {E.~H.}\ \bibnamefont {van Nes}}, \bibinfo {author} {\bibfnamefont {M.}~\bibnamefont {Rietkerk}}, \ and\ \bibinfo {author} {\bibfnamefont {G.}~\bibnamefont {Sugihara}},\ }\bibfield  {title} {\enquote {\bibinfo {title} {Early warning signals for critical transitions},}\ }\href@noop {} {\bibfield  {journal} {\bibinfo  {journal} {Nature}\ }\textbf {\bibinfo {volume} {461}},\ \bibinfo {pages} {53--59} (\bibinfo {year} {2009})}\BibitemShut {NoStop}%
\bibitem [{\citenamefont {Lenton}(2013)}]{2013Lenton}%
  \BibitemOpen
  \bibfield  {author} {\bibinfo {author} {\bibfnamefont {T.~M.}\ \bibnamefont {Lenton}},\ }\bibfield  {title} {\enquote {\bibinfo {title} {Environmental tipping points},}\ }\href {\doibase 10.1146/annurev-environ-102511-084654} {\bibfield  {journal} {\bibinfo  {journal} {Annual Review of Environment and Resources}\ }\textbf {\bibinfo {volume} {38}},\ \bibinfo {pages} {1--29} (\bibinfo {year} {2013})}\BibitemShut {NoStop}%
\bibitem [{\citenamefont {Brummitt}, \citenamefont {Barnett},\ and\ \citenamefont {D'Souza}(2015{\natexlab{a}})}]{2015BrummittBarnettDSouza}%
  \BibitemOpen
  \bibfield  {author} {\bibinfo {author} {\bibfnamefont {C.~D.}\ \bibnamefont {Brummitt}}, \bibinfo {author} {\bibfnamefont {G.}~\bibnamefont {Barnett}}, \ and\ \bibinfo {author} {\bibfnamefont {R.~M.}\ \bibnamefont {D'Souza}},\ }\bibfield  {title} {\enquote {\bibinfo {title} {Coupled catastrophes: sudden shifts cascade and hop among interdependent systems},}\ }\href {\doibase 10.1098/rsif.2015.0712} {\bibfield  {journal} {\bibinfo  {journal} {Journal of the Royal Society Interface}\ }\textbf {\bibinfo {volume} {12}},\ \bibinfo {pages} {0712} (\bibinfo {year} {2015}{\natexlab{a}})}\BibitemShut {NoStop}%
\bibitem [{\citenamefont {Feudel}, \citenamefont {Pisarchik},\ and\ \citenamefont {Showalter}(2018)}]{2018FeudelPisarchikShowalter}%
  \BibitemOpen
  \bibfield  {author} {\bibinfo {author} {\bibfnamefont {U.}~\bibnamefont {Feudel}}, \bibinfo {author} {\bibfnamefont {A.~N.}\ \bibnamefont {Pisarchik}}, \ and\ \bibinfo {author} {\bibfnamefont {K.}~\bibnamefont {Showalter}},\ }\bibfield  {title} {\enquote {\bibinfo {title} {Multistability and tipping: From mathematics and physics to climate and brain-minireview and preface to the focus issue},}\ }\href {\doibase 10.1063/1.5027718} {\bibfield  {journal} {\bibinfo  {journal} {Chaos}\ }\textbf {\bibinfo {volume} {28}},\ \bibinfo {pages} {033501} (\bibinfo {year} {2018})}\BibitemShut {NoStop}%
\bibitem [{\citenamefont {Datseris}, \citenamefont {Rossi},\ and\ \citenamefont {Wagemakers}(2023)}]{2023DatserisRossiWagemakers}%
  \BibitemOpen
  \bibfield  {author} {\bibinfo {author} {\bibfnamefont {G.}~\bibnamefont {Datseris}}, \bibinfo {author} {\bibfnamefont {K.~L.}\ \bibnamefont {Rossi}}, \ and\ \bibinfo {author} {\bibfnamefont {A.}~\bibnamefont {Wagemakers}},\ }\bibfield  {title} {\enquote {\bibinfo {title} {Framework for global stability analysis of dynamical systems},}\ }\href {\doibase 10.1063/5.0159675} {\bibfield  {journal} {\bibinfo  {journal} {Chaos}\ }\textbf {\bibinfo {volume} {33}},\ \bibinfo {pages} {073151} (\bibinfo {year} {2023})}\BibitemShut {NoStop}%
\bibitem [{\citenamefont {Caine}(1980)}]{1980Caine}%
  \BibitemOpen
  \bibfield  {author} {\bibinfo {author} {\bibfnamefont {N.}~\bibnamefont {Caine}},\ }\bibfield  {title} {\enquote {\bibinfo {title} {The rainfall intensity: Duration control of shallow landslides and debris flows},}\ }\href {\doibase 10.2307/520449} {\bibfield  {journal} {\bibinfo  {journal} {Geografiska Annaler. Series A, Physical Geography}\ }\textbf {\bibinfo {volume} {62}},\ \bibinfo {pages} {23--27} (\bibinfo {year} {1980})}\BibitemShut {NoStop}%
\bibitem [{\citenamefont {Larsen}\ and\ \citenamefont {Simon}(1993)}]{1993LarsenSimon}%
  \BibitemOpen
  \bibfield  {author} {\bibinfo {author} {\bibfnamefont {M.~C.}\ \bibnamefont {Larsen}}\ and\ \bibinfo {author} {\bibfnamefont {A.}~\bibnamefont {Simon}},\ }\bibfield  {title} {\enquote {\bibinfo {title} {A rainfall intensity-duration threshold for landslides in a humid-tropical environment, puerto rico},}\ }\href {\doibase 10.2307/521049} {\bibfield  {journal} {\bibinfo  {journal} {Geografiska Annaler. Series A, Physical Geography}\ }\textbf {\bibinfo {volume} {75}},\ \bibinfo {pages} {13--23} (\bibinfo {year} {1993})}\BibitemShut {NoStop}%
\bibitem [{\citenamefont {Chae}\ \emph {et~al.}(2017)\citenamefont {Chae}, \citenamefont {Park}, \citenamefont {Catani}, \citenamefont {Simoni},\ and\ \citenamefont {Berti}}]{2017ChaeParkCataniSimoniBerti}%
  \BibitemOpen
  \bibfield  {author} {\bibinfo {author} {\bibfnamefont {B.-G.}\ \bibnamefont {Chae}}, \bibinfo {author} {\bibfnamefont {H.-J.}\ \bibnamefont {Park}}, \bibinfo {author} {\bibfnamefont {F.}~\bibnamefont {Catani}}, \bibinfo {author} {\bibfnamefont {A.}~\bibnamefont {Simoni}}, \ and\ \bibinfo {author} {\bibfnamefont {M.}~\bibnamefont {Berti}},\ }\bibfield  {title} {\enquote {\bibinfo {title} {Landslide prediction, monitoring and early warning: a concise review of state-of-the-art},}\ }\href {\doibase 10.1007/s12303-017-0034-4} {\bibfield  {journal} {\bibinfo  {journal} {Geosciences Journal}\ }\textbf {\bibinfo {volume} {21}},\ \bibinfo {pages} {1033--1070} (\bibinfo {year} {2017})}\BibitemShut {NoStop}%
\bibitem [{\citenamefont {May}, \citenamefont {Levin},\ and\ \citenamefont {Sugihara}(2008)}]{2008MayLevinSugihara}%
  \BibitemOpen
  \bibfield  {author} {\bibinfo {author} {\bibfnamefont {R.~M.}\ \bibnamefont {May}}, \bibinfo {author} {\bibfnamefont {S.~A.}\ \bibnamefont {Levin}}, \ and\ \bibinfo {author} {\bibfnamefont {G.}~\bibnamefont {Sugihara}},\ }\bibfield  {title} {\enquote {\bibinfo {title} {Ecology for bankers},}\ }\href {\doibase 10.1038/451893a} {\bibfield  {journal} {\bibinfo  {journal} {Nature}\ }\textbf {\bibinfo {volume} {451}},\ \bibinfo {pages} {893--894} (\bibinfo {year} {2008})}\BibitemShut {NoStop}%
\bibitem [{\citenamefont {Gross}(2021)}]{2021Gross}%
  \BibitemOpen
  \bibfield  {author} {\bibinfo {author} {\bibfnamefont {T.}~\bibnamefont {Gross}},\ }\bibfield  {title} {\enquote {\bibinfo {title} {Not one, but many critical states: A dynamical systems perspective},}\ }\href {\doibase 10.3389/fncir.2021.614268} {\bibfield  {journal} {\bibinfo  {journal} {Frontiers in Neural Circuits}\ }\textbf {\bibinfo {volume} {15}},\ \bibinfo {pages} {614268} (\bibinfo {year} {2021})}\BibitemShut {NoStop}%
\bibitem [{\citenamefont {Ritchie}\ \emph {et~al.}(2021)\citenamefont {Ritchie}, \citenamefont {Clarke}, \citenamefont {Cox},\ and\ \citenamefont {Huntingford}}]{Ritchie2}%
  \BibitemOpen
  \bibfield  {author} {\bibinfo {author} {\bibfnamefont {P.}~\bibnamefont {Ritchie}}, \bibinfo {author} {\bibfnamefont {J.}~\bibnamefont {Clarke}}, \bibinfo {author} {\bibfnamefont {P.}~\bibnamefont {Cox}}, \ and\ \bibinfo {author} {\bibfnamefont {C.}~\bibnamefont {Huntingford}},\ }\bibfield  {title} {\enquote {\bibinfo {title} {Overshooting tipping point thresholds in a changing climate},}\ }\href {\doibase 10.1038/s41586-021-03263-2} {\bibfield  {journal} {\bibinfo  {journal} {Nature}\ }\textbf {\bibinfo {volume} {592}},\ \bibinfo {pages} {517--523} (\bibinfo {year} {2021})}\BibitemShut {NoStop}%
\bibitem [{\citenamefont {Wunderling}\ \emph {et~al.}(2023)\citenamefont {Wunderling}, \citenamefont {Winkelmann}, \citenamefont {Rockstr{\"o}m}, \citenamefont {Loriani}, \citenamefont {Armstrong~McKay}, \citenamefont {Ritchie}, \citenamefont {Sakschewski},\ and\ \citenamefont {Donges}}]{wunderling2023global}%
  \BibitemOpen
  \bibfield  {author} {\bibinfo {author} {\bibfnamefont {N.}~\bibnamefont {Wunderling}}, \bibinfo {author} {\bibfnamefont {R.}~\bibnamefont {Winkelmann}}, \bibinfo {author} {\bibfnamefont {J.}~\bibnamefont {Rockstr{\"o}m}}, \bibinfo {author} {\bibfnamefont {S.}~\bibnamefont {Loriani}}, \bibinfo {author} {\bibfnamefont {D.~I.}\ \bibnamefont {Armstrong~McKay}}, \bibinfo {author} {\bibfnamefont {P.~D.}\ \bibnamefont {Ritchie}}, \bibinfo {author} {\bibfnamefont {B.}~\bibnamefont {Sakschewski}}, \ and\ \bibinfo {author} {\bibfnamefont {J.~F.}\ \bibnamefont {Donges}},\ }\bibfield  {title} {\enquote {\bibinfo {title} {Global warming overshoots increase risks of climate tipping cascades in a network model},}\ }\href@noop {} {\bibfield  {journal} {\bibinfo  {journal} {Nature Climate Change}\ }\textbf {\bibinfo {volume} {13}},\ \bibinfo {pages} {75--82} (\bibinfo {year} {2023})}\BibitemShut {NoStop}%
\bibitem [{\citenamefont {Ritchie}, \citenamefont {Karabacak},\ and\ \citenamefont {Sieber}(2019)}]{Ritchie}%
  \BibitemOpen
  \bibfield  {author} {\bibinfo {author} {\bibfnamefont {P.}~\bibnamefont {Ritchie}}, \bibinfo {author} {\bibfnamefont {{\"O}.}~\bibnamefont {Karabacak}}, \ and\ \bibinfo {author} {\bibfnamefont {J.}~\bibnamefont {Sieber}},\ }\bibfield  {title} {\enquote {\bibinfo {title} {Inverse-square law between time and amplitude for crossing tipping thresholds},}\ }\href@noop {} {\bibfield  {journal} {\bibinfo  {journal} {Proc. R. Soc. A}\ }\textbf {\bibinfo {volume} {475}} (\bibinfo {year} {2019})}\BibitemShut {NoStop}%
\bibitem [{\citenamefont {Kuehn}(2011)}]{2011Kuehn}%
  \BibitemOpen
  \bibfield  {author} {\bibinfo {author} {\bibfnamefont {C.}~\bibnamefont {Kuehn}},\ }\bibfield  {title} {\enquote {\bibinfo {title} {A mathematical framework for critical transitions: Bifurcations, fast-slow systems and stochastic dynamics},}\ }\href {\doibase 10.1016/j.physd.2011.02.012} {\bibfield  {journal} {\bibinfo  {journal} {Physica D: Nonlinear Phenomena}\ }\textbf {\bibinfo {volume} {240}},\ \bibinfo {pages} {1020--1035} (\bibinfo {year} {2011})}\BibitemShut {NoStop}%
\bibitem [{\citenamefont {Ashwin}\ \emph {et~al.}(2012)\citenamefont {Ashwin}, \citenamefont {Wieczorek}, \citenamefont {Vitolo},\ and\ \citenamefont {Cox}}]{2012AshwinWieczorekVitoloCox}%
  \BibitemOpen
  \bibfield  {author} {\bibinfo {author} {\bibfnamefont {P.}~\bibnamefont {Ashwin}}, \bibinfo {author} {\bibfnamefont {S.}~\bibnamefont {Wieczorek}}, \bibinfo {author} {\bibfnamefont {R.}~\bibnamefont {Vitolo}}, \ and\ \bibinfo {author} {\bibfnamefont {P.}~\bibnamefont {Cox}},\ }\bibfield  {title} {\enquote {\bibinfo {title} {Tipping points in open systems: bifurcation, noise-induced and rate-dependent examples in the climate system.}}\ }\href@noop {} {\bibfield  {journal} {\bibinfo  {journal} {Philos. Trans. R. Soc. A Math. Phys. Eng. Sci.}\ }\textbf {\bibinfo {volume} {370}},\ \bibinfo {pages} {1166--1184} (\bibinfo {year} {2012})}\BibitemShut {NoStop}%
\bibitem [{\citenamefont {Brummitt}, \citenamefont {Barnett},\ and\ \citenamefont {D'Souza}(2015{\natexlab{b}})}]{brummitt2015coupled}%
  \BibitemOpen
  \bibfield  {author} {\bibinfo {author} {\bibfnamefont {C.~D.}\ \bibnamefont {Brummitt}}, \bibinfo {author} {\bibfnamefont {G.}~\bibnamefont {Barnett}}, \ and\ \bibinfo {author} {\bibfnamefont {R.~M.}\ \bibnamefont {D'Souza}},\ }\bibfield  {title} {\enquote {\bibinfo {title} {Coupled catastrophes: sudden shifts cascade and hop among interdependent systems},}\ }\href@noop {} {\bibfield  {journal} {\bibinfo  {journal} {Journal of The Royal Society Interface}\ }\textbf {\bibinfo {volume} {12}},\ \bibinfo {pages} {20150712} (\bibinfo {year} {2015}{\natexlab{b}})}\BibitemShut {NoStop}%
\bibitem [{\citenamefont {Rocha}\ \emph {et~al.}(2018)\citenamefont {Rocha}, \citenamefont {Peterson}, \citenamefont {Bodin},\ and\ \citenamefont {Levin}}]{rocha2018cascading}%
  \BibitemOpen
  \bibfield  {author} {\bibinfo {author} {\bibfnamefont {J.~C.}\ \bibnamefont {Rocha}}, \bibinfo {author} {\bibfnamefont {G.}~\bibnamefont {Peterson}}, \bibinfo {author} {\bibfnamefont {{\"O}.}~\bibnamefont {Bodin}}, \ and\ \bibinfo {author} {\bibfnamefont {S.}~\bibnamefont {Levin}},\ }\bibfield  {title} {\enquote {\bibinfo {title} {Cascading regime shifts within and across scales},}\ }\href@noop {} {\bibfield  {journal} {\bibinfo  {journal} {Science}\ }\textbf {\bibinfo {volume} {362}},\ \bibinfo {pages} {1379--1383} (\bibinfo {year} {2018})}\BibitemShut {NoStop}%
\bibitem [{\citenamefont {Drouet}\ \emph {et~al.}(2021)\citenamefont {Drouet}, \citenamefont {Bosetti}, \citenamefont {Padoan}, \citenamefont {Aleluia~Reis}, \citenamefont {Bertram}, \citenamefont {Dalla~Longa}, \citenamefont {Despr{\'e}s}, \citenamefont {Emmerling}, \citenamefont {Fosse}, \citenamefont {Fragkiadakis} \emph {et~al.}}]{drouet2021net}%
  \BibitemOpen
  \bibfield  {author} {\bibinfo {author} {\bibfnamefont {L.}~\bibnamefont {Drouet}}, \bibinfo {author} {\bibfnamefont {V.}~\bibnamefont {Bosetti}}, \bibinfo {author} {\bibfnamefont {S.~A.}\ \bibnamefont {Padoan}}, \bibinfo {author} {\bibfnamefont {L.}~\bibnamefont {Aleluia~Reis}}, \bibinfo {author} {\bibfnamefont {C.}~\bibnamefont {Bertram}}, \bibinfo {author} {\bibfnamefont {F.}~\bibnamefont {Dalla~Longa}}, \bibinfo {author} {\bibfnamefont {J.}~\bibnamefont {Despr{\'e}s}}, \bibinfo {author} {\bibfnamefont {J.}~\bibnamefont {Emmerling}}, \bibinfo {author} {\bibfnamefont {F.}~\bibnamefont {Fosse}}, \bibinfo {author} {\bibfnamefont {K.}~\bibnamefont {Fragkiadakis}},  \emph {et~al.},\ }\bibfield  {title} {\enquote {\bibinfo {title} {Net zero-emission pathways reduce the physical and economic risks of climate change},}\ }\href@noop {} {\bibfield  {journal} {\bibinfo  {journal} {Nature Climate Change}\ }\textbf {\bibinfo {volume} {11}},\ \bibinfo {pages} {1070--1076} (\bibinfo {year} {2021})}\BibitemShut {NoStop}%
\bibitem [{\citenamefont {Riahi}\ \emph {et~al.}(2021)\citenamefont {Riahi}, \citenamefont {Bertram}, \citenamefont {Huppmann}, \citenamefont {Rogelj}, \citenamefont {Bosetti}, \citenamefont {Cabardos}, \citenamefont {Deppermann}, \citenamefont {Drouet}, \citenamefont {Frank}, \citenamefont {Fricko} \emph {et~al.}}]{riahi2021cost}%
  \BibitemOpen
  \bibfield  {author} {\bibinfo {author} {\bibfnamefont {K.}~\bibnamefont {Riahi}}, \bibinfo {author} {\bibfnamefont {C.}~\bibnamefont {Bertram}}, \bibinfo {author} {\bibfnamefont {D.}~\bibnamefont {Huppmann}}, \bibinfo {author} {\bibfnamefont {J.}~\bibnamefont {Rogelj}}, \bibinfo {author} {\bibfnamefont {V.}~\bibnamefont {Bosetti}}, \bibinfo {author} {\bibfnamefont {A.-M.}\ \bibnamefont {Cabardos}}, \bibinfo {author} {\bibfnamefont {A.}~\bibnamefont {Deppermann}}, \bibinfo {author} {\bibfnamefont {L.}~\bibnamefont {Drouet}}, \bibinfo {author} {\bibfnamefont {S.}~\bibnamefont {Frank}}, \bibinfo {author} {\bibfnamefont {O.}~\bibnamefont {Fricko}},  \emph {et~al.},\ }\bibfield  {title} {\enquote {\bibinfo {title} {Cost and attainability of meeting stringent climate targets without overshoot},}\ }\href@noop {} {\bibfield  {journal} {\bibinfo  {journal} {Nature Climate Change}\ }\textbf {\bibinfo {volume} {11}},\ \bibinfo {pages} {1063--1069} (\bibinfo {year} {2021})}\BibitemShut {NoStop}%
\bibitem [{\citenamefont {Rogelj}\ \emph {et~al.}(2019)\citenamefont {Rogelj}, \citenamefont {Huppmann}, \citenamefont {Krey}, \citenamefont {Riahi}, \citenamefont {Clarke}, \citenamefont {Gidden}, \citenamefont {Nicholls},\ and\ \citenamefont {Meinshausen}}]{rogelj2019new}%
  \BibitemOpen
  \bibfield  {author} {\bibinfo {author} {\bibfnamefont {J.}~\bibnamefont {Rogelj}}, \bibinfo {author} {\bibfnamefont {D.}~\bibnamefont {Huppmann}}, \bibinfo {author} {\bibfnamefont {V.}~\bibnamefont {Krey}}, \bibinfo {author} {\bibfnamefont {K.}~\bibnamefont {Riahi}}, \bibinfo {author} {\bibfnamefont {L.}~\bibnamefont {Clarke}}, \bibinfo {author} {\bibfnamefont {M.}~\bibnamefont {Gidden}}, \bibinfo {author} {\bibfnamefont {Z.}~\bibnamefont {Nicholls}}, \ and\ \bibinfo {author} {\bibfnamefont {M.}~\bibnamefont {Meinshausen}},\ }\bibfield  {title} {\enquote {\bibinfo {title} {A new scenario logic for the {Paris Agreement} long-term temperature goal},}\ }\href@noop {} {\bibfield  {journal} {\bibinfo  {journal} {Nature}\ }\textbf {\bibinfo {volume} {573}},\ \bibinfo {pages} {357--363} (\bibinfo {year} {2019})}\BibitemShut {NoStop}%
\bibitem [{\citenamefont {Bender}\ and\ \citenamefont {Orszag}(1987)}]{1987BenderOrszag-book}%
  \BibitemOpen
  \bibfield  {author} {\bibinfo {author} {\bibfnamefont {C.~M.}\ \bibnamefont {Bender}}\ and\ \bibinfo {author} {\bibfnamefont {S.~A.}\ \bibnamefont {Orszag}},\ }\href@noop {} {\emph {\bibinfo {title} {Advanced Mathematical Methods for Scientists and Engineers}}},\ Mathematics Series\ (\bibinfo  {publisher} {McGraw-Hill},\ \bibinfo {address} {Singapore},\ \bibinfo {year} {1987})\BibitemShut {NoStop}%
\bibitem [{\citenamefont {Guckenheimer}\ and\ \citenamefont {Holmes}(1983)}]{1983GuckenheimerHolmes}%
  \BibitemOpen
  \bibfield  {author} {\bibinfo {author} {\bibfnamefont {J.}~\bibnamefont {Guckenheimer}}\ and\ \bibinfo {author} {\bibfnamefont {P.~J.}\ \bibnamefont {Holmes}},\ }\href@noop {} {\emph {\bibinfo {title} {Nonlinear Oscillations, Dynamical Systems, and Bifurcations of Vector Fields}}}\ (\bibinfo  {publisher} {Springer},\ \bibinfo {year} {1983})\BibitemShut {NoStop}%
\bibitem [{\citenamefont {Lenton}\ \emph {et~al.}(2008)\citenamefont {Lenton}, \citenamefont {Held}, \citenamefont {Kriegler}, \citenamefont {Hall}, \citenamefont {Lucht}, \citenamefont {Rahmstorf},\ and\ \citenamefont {Schellnhuber}}]{tippingbasics}%
  \BibitemOpen
  \bibfield  {author} {\bibinfo {author} {\bibfnamefont {T.~M.}\ \bibnamefont {Lenton}}, \bibinfo {author} {\bibfnamefont {H.}~\bibnamefont {Held}}, \bibinfo {author} {\bibfnamefont {E.}~\bibnamefont {Kriegler}}, \bibinfo {author} {\bibfnamefont {J.~W.}\ \bibnamefont {Hall}}, \bibinfo {author} {\bibfnamefont {W.}~\bibnamefont {Lucht}}, \bibinfo {author} {\bibfnamefont {S.}~\bibnamefont {Rahmstorf}}, \ and\ \bibinfo {author} {\bibfnamefont {H.~J.}\ \bibnamefont {Schellnhuber}},\ }\bibfield  {title} {\enquote {\bibinfo {title} {Tipping elements in the earth's climate system},}\ }\href {\doibase 10.1073/pnas.0705414105} {\bibfield  {journal} {\bibinfo  {journal} {Proceedings of the National Academy of Sciences}\ }\textbf {\bibinfo {volume} {105}},\ \bibinfo {pages} {1786--1793} (\bibinfo {year} {2008})}\BibitemShut {NoStop}%
\bibitem [{\citenamefont {Armstrong~McKay}\ \emph {et~al.}(2022)\citenamefont {Armstrong~McKay}, \citenamefont {Staal}, \citenamefont {Abrams}, \citenamefont {Winkelmann}, \citenamefont {Sakschewski}, \citenamefont {Loriani}, \citenamefont {Fetzer}, \citenamefont {Cornell}, \citenamefont {Rockstr{\"o}m},\ and\ \citenamefont {Lenton}}]{armstrong2022exceeding}%
  \BibitemOpen
  \bibfield  {author} {\bibinfo {author} {\bibfnamefont {D.~I.}\ \bibnamefont {Armstrong~McKay}}, \bibinfo {author} {\bibfnamefont {A.}~\bibnamefont {Staal}}, \bibinfo {author} {\bibfnamefont {J.~F.}\ \bibnamefont {Abrams}}, \bibinfo {author} {\bibfnamefont {R.}~\bibnamefont {Winkelmann}}, \bibinfo {author} {\bibfnamefont {B.}~\bibnamefont {Sakschewski}}, \bibinfo {author} {\bibfnamefont {S.}~\bibnamefont {Loriani}}, \bibinfo {author} {\bibfnamefont {I.}~\bibnamefont {Fetzer}}, \bibinfo {author} {\bibfnamefont {S.~E.}\ \bibnamefont {Cornell}}, \bibinfo {author} {\bibfnamefont {J.}~\bibnamefont {Rockstr{\"o}m}}, \ and\ \bibinfo {author} {\bibfnamefont {T.~M.}\ \bibnamefont {Lenton}},\ }\bibfield  {title} {\enquote {\bibinfo {title} {Exceeding {$1.5^\circ$C} global warming could trigger multiple climate tipping points},}\ }\href@noop {} {\bibfield  {journal} {\bibinfo  {journal} {Science}\ }\textbf {\bibinfo {volume} {377}},\ \bibinfo {pages} {eabn7950} (\bibinfo {year} {2022})}\BibitemShut {NoStop}%
\bibitem [{\citenamefont {Staal}\ \emph {et~al.}(2015)\citenamefont {Staal}, \citenamefont {Dekker}, \citenamefont {Hirota},\ and\ \citenamefont {van Nes}}]{staal2015synergistic}%
  \BibitemOpen
  \bibfield  {author} {\bibinfo {author} {\bibfnamefont {A.}~\bibnamefont {Staal}}, \bibinfo {author} {\bibfnamefont {S.~C.}\ \bibnamefont {Dekker}}, \bibinfo {author} {\bibfnamefont {M.}~\bibnamefont {Hirota}}, \ and\ \bibinfo {author} {\bibfnamefont {E.~H.}\ \bibnamefont {van Nes}},\ }\bibfield  {title} {\enquote {\bibinfo {title} {Synergistic effects of drought and deforestation on the resilience of the south-eastern {Amazon} rainforest},}\ }\href@noop {} {\bibfield  {journal} {\bibinfo  {journal} {Ecological Complexity}\ }\textbf {\bibinfo {volume} {22}},\ \bibinfo {pages} {65--75} (\bibinfo {year} {2015})}\BibitemShut {NoStop}%
\bibitem [{\citenamefont {Levermann}\ and\ \citenamefont {Winkelmann}(2016)}]{levermann2016simple}%
  \BibitemOpen
  \bibfield  {author} {\bibinfo {author} {\bibfnamefont {A.}~\bibnamefont {Levermann}}\ and\ \bibinfo {author} {\bibfnamefont {R.}~\bibnamefont {Winkelmann}},\ }\bibfield  {title} {\enquote {\bibinfo {title} {A simple equation for the melt elevation feedback of ice sheets},}\ }\href@noop {} {\bibfield  {journal} {\bibinfo  {journal} {The Cryosphere}\ }\textbf {\bibinfo {volume} {10}},\ \bibinfo {pages} {1799--1807} (\bibinfo {year} {2016})}\BibitemShut {NoStop}%
\bibitem [{\citenamefont {Stommel}(1961)}]{stommel1961thermohaline}%
  \BibitemOpen
  \bibfield  {author} {\bibinfo {author} {\bibfnamefont {H.}~\bibnamefont {Stommel}},\ }\bibfield  {title} {\enquote {\bibinfo {title} {Thermohaline convection with two stable regimes of flow},}\ }\href@noop {} {\bibfield  {journal} {\bibinfo  {journal} {Tellus}\ }\textbf {\bibinfo {volume} {13}},\ \bibinfo {pages} {224--230} (\bibinfo {year} {1961})}\BibitemShut {NoStop}%
\bibitem [{\citenamefont {Cessi}(1994)}]{cessi1994simple}%
  \BibitemOpen
  \bibfield  {author} {\bibinfo {author} {\bibfnamefont {P.}~\bibnamefont {Cessi}},\ }\bibfield  {title} {\enquote {\bibinfo {title} {A simple box model of stochastically forced thermohaline flow},}\ }\href@noop {} {\bibfield  {journal} {\bibinfo  {journal} {Journal of Physical Oceanography}\ }\textbf {\bibinfo {volume} {24}},\ \bibinfo {pages} {1911--1920} (\bibinfo {year} {1994})}\BibitemShut {NoStop}%
\bibitem [{\citenamefont {Wunderling}\ \emph {et~al.}(2021)\citenamefont {Wunderling}, \citenamefont {Kr{\"o}nke}, \citenamefont {Wohlfarth}, \citenamefont {Kohler}, \citenamefont {Heitzig}, \citenamefont {Staal}, \citenamefont {Willner}, \citenamefont {Winkelmann},\ and\ \citenamefont {Donges}}]{wunderling2021modelling}%
  \BibitemOpen
  \bibfield  {author} {\bibinfo {author} {\bibfnamefont {N.}~\bibnamefont {Wunderling}}, \bibinfo {author} {\bibfnamefont {J.}~\bibnamefont {Kr{\"o}nke}}, \bibinfo {author} {\bibfnamefont {V.}~\bibnamefont {Wohlfarth}}, \bibinfo {author} {\bibfnamefont {J.}~\bibnamefont {Kohler}}, \bibinfo {author} {\bibfnamefont {J.}~\bibnamefont {Heitzig}}, \bibinfo {author} {\bibfnamefont {A.}~\bibnamefont {Staal}}, \bibinfo {author} {\bibfnamefont {S.}~\bibnamefont {Willner}}, \bibinfo {author} {\bibfnamefont {R.}~\bibnamefont {Winkelmann}}, \ and\ \bibinfo {author} {\bibfnamefont {J.~F.}\ \bibnamefont {Donges}},\ }\bibfield  {title} {\enquote {\bibinfo {title} {Modelling nonlinear dynamics of interacting tipping elements on complex networks: the pycascades package},}\ }\href@noop {} {\bibfield  {journal} {\bibinfo  {journal} {The European Physical Journal Special Topics}\ }\textbf {\bibinfo {volume} {230}},\ \bibinfo {pages} {3163--3176} (\bibinfo {year} {2021})}\BibitemShut {NoStop}%
\bibitem [{\citenamefont {Kohler}\ \emph {et~al.}(2021)\citenamefont {Kohler}, \citenamefont {Wunderling}, \citenamefont {Donges},\ and\ \citenamefont {Vollmer}}]{2021KohlerWunderlingDongesVollmer}%
  \BibitemOpen
  \bibfield  {author} {\bibinfo {author} {\bibfnamefont {J.}~\bibnamefont {Kohler}}, \bibinfo {author} {\bibfnamefont {N.}~\bibnamefont {Wunderling}}, \bibinfo {author} {\bibfnamefont {J.~F.}\ \bibnamefont {Donges}}, \ and\ \bibinfo {author} {\bibfnamefont {J.}~\bibnamefont {Vollmer}},\ }\bibfield  {title} {\enquote {\bibinfo {title} {Complex networks of interacting stochastic tipping elements: cooperativity of phase separation in the large-system limit},}\ }\href {\doibase 10.1103/PhysRevE.104.044301} {\bibfield  {journal} {\bibinfo  {journal} {Physical Review E}\ }\textbf {\bibinfo {volume} {104}},\ \bibinfo {pages} {044301} (\bibinfo {year} {2021})},\ \Eprint {http://arxiv.org/abs/2104.09299v1} {2104.09299v1} \BibitemShut {NoStop}%
\bibitem [{\citenamefont {Masson-Delmotte}\ \emph {et~al.}(2018)\citenamefont {Masson-Delmotte}, \citenamefont {Zhai}, \citenamefont {P{\"o}rtner}, \citenamefont {Roberts}, \citenamefont {Skea}, \citenamefont {Shukla}, \citenamefont {Pirani}, \citenamefont {Moufouma-Okia}, \citenamefont {P{\'e}an}, \citenamefont {Pidcock} \emph {et~al.}}]{masson2018global}%
  \BibitemOpen
  \bibfield  {author} {\bibinfo {author} {\bibfnamefont {V.}~\bibnamefont {Masson-Delmotte}}, \bibinfo {author} {\bibfnamefont {P.}~\bibnamefont {Zhai}}, \bibinfo {author} {\bibfnamefont {H.-O.}\ \bibnamefont {P{\"o}rtner}}, \bibinfo {author} {\bibfnamefont {D.}~\bibnamefont {Roberts}}, \bibinfo {author} {\bibfnamefont {J.}~\bibnamefont {Skea}}, \bibinfo {author} {\bibfnamefont {P.~R.}\ \bibnamefont {Shukla}}, \bibinfo {author} {\bibfnamefont {A.}~\bibnamefont {Pirani}}, \bibinfo {author} {\bibfnamefont {W.}~\bibnamefont {Moufouma-Okia}}, \bibinfo {author} {\bibfnamefont {C.}~\bibnamefont {P{\'e}an}}, \bibinfo {author} {\bibfnamefont {R.}~\bibnamefont {Pidcock}},  \emph {et~al.},\ }\bibfield  {title} {\enquote {\bibinfo {title} {An {IPCC} special report on the impacts of global warming of {$1.5^\circ$C}},}\ }\href@noop {} {\bibfield  {journal} {\bibinfo  {journal} {IPCC}\ }\textbf {\bibinfo {volume} {1}},\ \bibinfo {pages} {43--50} (\bibinfo {year} {2018})}\BibitemShut {NoStop}%
\bibitem [{\citenamefont {Schleussner}\ \emph {et~al.}(2022)\citenamefont {Schleussner}, \citenamefont {Ganti}, \citenamefont {Rogelj},\ and\ \citenamefont {Gidden}}]{schleussner2022emission}%
  \BibitemOpen
  \bibfield  {author} {\bibinfo {author} {\bibfnamefont {C.-F.}\ \bibnamefont {Schleussner}}, \bibinfo {author} {\bibfnamefont {G.}~\bibnamefont {Ganti}}, \bibinfo {author} {\bibfnamefont {J.}~\bibnamefont {Rogelj}}, \ and\ \bibinfo {author} {\bibfnamefont {M.~J.}\ \bibnamefont {Gidden}},\ }\bibfield  {title} {\enquote {\bibinfo {title} {An emission pathway classification reflecting the {Paris Agreement} climate objectives},}\ }\href@noop {} {\bibfield  {journal} {\bibinfo  {journal} {Communications Earth \& Environment}\ }\textbf {\bibinfo {volume} {3}},\ \bibinfo {pages} {135} (\bibinfo {year} {2022})}\BibitemShut {NoStop}%
\bibitem [{\citenamefont {Rogelj}\ \emph {et~al.}(2023)\citenamefont {Rogelj}, \citenamefont {Fransen}, \citenamefont {den Elzen}, \citenamefont {Lamboll}, \citenamefont {Schumer}, \citenamefont {Kuramochi}, \citenamefont {Hans}, \citenamefont {Mooldijk},\ and\ \citenamefont {Portugal-Pereira}}]{rogelj2023credibility-gap}%
  \BibitemOpen
  \bibfield  {author} {\bibinfo {author} {\bibfnamefont {J.}~\bibnamefont {Rogelj}}, \bibinfo {author} {\bibfnamefont {T.}~\bibnamefont {Fransen}}, \bibinfo {author} {\bibfnamefont {M.}~\bibnamefont {den Elzen}}, \bibinfo {author} {\bibfnamefont {R.}~\bibnamefont {Lamboll}}, \bibinfo {author} {\bibfnamefont {C.}~\bibnamefont {Schumer}}, \bibinfo {author} {\bibfnamefont {T.}~\bibnamefont {Kuramochi}}, \bibinfo {author} {\bibfnamefont {F.}~\bibnamefont {Hans}}, \bibinfo {author} {\bibfnamefont {S.}~\bibnamefont {Mooldijk}}, \ and\ \bibinfo {author} {\bibnamefont {Portugal-Pereira}},\ }\bibfield  {title} {\enquote {\bibinfo {title} {Credibility gap in net-zero climate targets leaves world at high risk},}\ }\href@noop {} {\bibfield  {journal} {\bibinfo  {journal} {Science}\ }\textbf {\bibinfo {volume} {380}},\ \bibinfo {pages} {1014–1016} (\bibinfo {year} {2023})}\BibitemShut {NoStop}%
\bibitem [{\citenamefont {Masson-Delmotte}\ \emph {et~al.}(2021)\citenamefont {Masson-Delmotte}, \citenamefont {Zhai}, \citenamefont {Pirani}, \citenamefont {Connors}, \citenamefont {Péan}, \citenamefont {Berger}, \citenamefont {Caud}, \citenamefont {Chen}, \citenamefont {Goldfarb}, \citenamefont {Gomis}, \citenamefont {Huang}, \citenamefont {Leitzell}, \citenamefont {Lonnoy}, \citenamefont {Matthews}, \citenamefont {Maycock}, \citenamefont {Waterfield}, \citenamefont {Yelekçi}, \citenamefont {Yu},\ and\ \citenamefont {Zhou}}]{masson2021ipcc}%
  \BibitemOpen
  \bibinfo {editor} {\bibfnamefont {V.}~\bibnamefont {Masson-Delmotte}}, \bibinfo {editor} {\bibfnamefont {P.}~\bibnamefont {Zhai}}, \bibinfo {editor} {\bibfnamefont {A.}~\bibnamefont {Pirani}}, \bibinfo {editor} {\bibfnamefont {S.}~\bibnamefont {Connors}}, \bibinfo {editor} {\bibfnamefont {C.}~\bibnamefont {Péan}}, \bibinfo {editor} {\bibfnamefont {S.}~\bibnamefont {Berger}}, \bibinfo {editor} {\bibfnamefont {N.}~\bibnamefont {Caud}}, \bibinfo {editor} {\bibfnamefont {Y.}~\bibnamefont {Chen}}, \bibinfo {editor} {\bibfnamefont {L.}~\bibnamefont {Goldfarb}}, \bibinfo {editor} {\bibfnamefont {M.}~\bibnamefont {Gomis}}, \bibinfo {editor} {\bibfnamefont {M.}~\bibnamefont {Huang}}, \bibinfo {editor} {\bibfnamefont {K.}~\bibnamefont {Leitzell}}, \bibinfo {editor} {\bibfnamefont {E.}~\bibnamefont {Lonnoy}}, \bibinfo {editor} {\bibfnamefont {J.}~\bibnamefont {Matthews}}, \bibinfo {editor} {\bibfnamefont {T.}~\bibnamefont {Maycock}}, \bibinfo {editor} {\bibfnamefont {T.}~\bibnamefont {Waterfield}}, \bibinfo {editor}
  {\bibfnamefont {O.}~\bibnamefont {Yelekçi}}, \bibinfo {editor} {\bibfnamefont {R.}~\bibnamefont {Yu}}, \ and\ \bibinfo {editor} {\bibfnamefont {B.}~\bibnamefont {Zhou}},\ eds.,\ \enquote {\bibinfo {title} {Climate change 2021: {T}he physical science basis. {C}ontribution of working group {I} to the sixth assessment report of the intergovernmental panel on climate change},}\ \ (\bibinfo  {publisher} {Cambridge University Press},\ \bibinfo {address} {Cambridge, United Kingdom and New York, NY, USA},\ \bibinfo {year} {2021})\ Chap.\ \bibinfo {chapter} {{IPCC}, 2021: {S}ummary for Policymakers}, pp.\ \bibinfo {pages} {3--32}\BibitemShut {NoStop}%
\bibitem [{\citenamefont {Kemp}\ \emph {et~al.}(2022)\citenamefont {Kemp}, \citenamefont {Xu}, \citenamefont {Depledge}, \citenamefont {Ebi}, \citenamefont {Gibbins}, \citenamefont {Kohler}, \citenamefont {Rockstr{\"o}m}, \citenamefont {Scheffer}, \citenamefont {Schellnhuber}, \citenamefont {Steffen} \emph {et~al.}}]{kemp2022climate}%
  \BibitemOpen
  \bibfield  {author} {\bibinfo {author} {\bibfnamefont {L.}~\bibnamefont {Kemp}}, \bibinfo {author} {\bibfnamefont {C.}~\bibnamefont {Xu}}, \bibinfo {author} {\bibfnamefont {J.}~\bibnamefont {Depledge}}, \bibinfo {author} {\bibfnamefont {K.~L.}\ \bibnamefont {Ebi}}, \bibinfo {author} {\bibfnamefont {G.}~\bibnamefont {Gibbins}}, \bibinfo {author} {\bibfnamefont {T.~A.}\ \bibnamefont {Kohler}}, \bibinfo {author} {\bibfnamefont {J.}~\bibnamefont {Rockstr{\"o}m}}, \bibinfo {author} {\bibfnamefont {M.}~\bibnamefont {Scheffer}}, \bibinfo {author} {\bibfnamefont {H.~J.}\ \bibnamefont {Schellnhuber}}, \bibinfo {author} {\bibfnamefont {W.}~\bibnamefont {Steffen}},  \emph {et~al.},\ }\bibfield  {title} {\enquote {\bibinfo {title} {Climate endgame: Exploring catastrophic climate change scenarios},}\ }\href@noop {} {\bibfield  {journal} {\bibinfo  {journal} {Proceedings of the National Academy of Sciences}\ }\textbf {\bibinfo {volume} {119}},\ \bibinfo {pages} {e2108146119} (\bibinfo {year} {2022})}\BibitemShut
  {NoStop}%
\end{thebibliography}%



\end{document}